\documentclass[aps, pra, reprint, superscriptaddress]{revtex4-2}
\usepackage{graphicx}
\usepackage{amsmath}
\usepackage{amsfonts}
\usepackage{amssymb}
\usepackage[mathscr]{euscript}
\usepackage{braket}
\usepackage{xcolor}
\usepackage{tikz}
\usepackage[caption=false]{subfig}
\usepackage{adjustbox}
\usetikzlibrary{quantikz2}
\usepackage[colorlinks=true]{hyperref}
\usepackage{orcidlink}
\usepackage{float}
\usepackage{enumitem}

\usepackage{silence}
\newcommand{\arrmark}{\push{\;\rightarrow\;}\qw}

\begin{document}
\title{\texorpdfstring{Benchmarking a machine-learning differential equations solver\\on a neutral-atom logical processor}{Benchmarking a machine-learning differential equations solver on a neutral-atom logical processor}}

\author{Pauline Mathiot}
\thanks{These authors contributed equally to this work.}
\author{Elio Garnaoui}
\thanks{These authors contributed equally to this work.}
\affiliation{PASQAL SAS, 24 rue Emile Baudot, 91120 Palaiseau, France}
\author{Axel-Ugo Leriche}
\thanks{These authors contributed equally to this work.}
\affiliation{PASQAL SAS, 24 rue Emile Baudot, 91120 Palaiseau, France}
\affiliation{Université Paris-Saclay, Institut d’Optique Graduate School,
CNRS, Laboratoire Charles Fabry, 91127 Palaiseau Cedex, France}
\author{Evan Philip}
\thanks{These authors contributed equally to this work.}
\author{Boris Albrecht}
\author{Clémence Briosne-Fréjaville}
\author{Lorenzo Cardarelli}
\author{Antoine Cornillot}
\author{Gwennolé Cournez}
\author{Luc Couturier}
\author{Julius De Hond}
\author{Rebecca El Koussaifi}
\author{Thomas Eritzpokoff}
\author{Florian Fasola}
\author{Antonio Andrea Gentile}
\author{Casper Gyurik}
\author{Clotilde Hamot}
\author{Loïc Henriet}
\author{Gaétan Hercé}
\author{Michael Kaicher}
\author{Lucas Lassablière}
\author{François-Marie Le Régent}
\author{Edgar Leroux}
\author{Yohann Machu}
\author{Hadriel Mamann}
\author{Luis Ortiz}
\author{Annie Paine}
\author{Thomas Pansiot}
\author{Arnaud Peloquin}
\author{Francisco Ponciano}
\author{Julien Ripoll}
\author{Raja Selvarajan}
\author{Adrien Signoles}
\author{Henrique Silvério}
\author{Siddhy Tan}
\author{Marie Taouzinet}
\author{Selim Touati}
\author{Louis Vignoli}
\affiliation{PASQAL SAS, 24 rue Emile Baudot, 91120 Palaiseau, France}
\author{Antoine Browaeys}
\affiliation{PASQAL SAS, 24 rue Emile Baudot, 91120 Palaiseau, France}
\affiliation{Université Paris-Saclay, Institut d’Optique Graduate School,
CNRS, Laboratoire Charles Fabry, 91127 Palaiseau Cedex, France}
\author{Pascal Scholl}
\email{pascal.scholl@pasqal.com}
\affiliation{PASQAL SAS, 24 rue Emile Baudot, 91120 Palaiseau, France}
\date{\today}

\begin{abstract}
We report on a performance comparison between physical and logical computations on a prototypical machine-learning application: solving differential equations using quantum kernel methods. The algorithm is implemented on an atom-based logical quantum processor, both at the physical and logical levels. 
We show that the kernel estimated from the logical implementation performs better than its physical counterpart on relevant metrics. 
We observe how such performance improvement can be traced back to specific noise-induced errors detected by the chosen encoding.
We apply the computed quantum kernel to the task of solving differential equations, confirming how the superior performance of a logical quantum kernel is retained also at an end-to-end applicative level. 
Our findings show that experimental validation of end-to-end protocols can already highlight the positive impact of fault-tolerant implementations despite their higher quantum resource count, and guide application-informed architectural choices.
\end{abstract}

\maketitle

\section{Introduction}

The recent advancement of quantum processors towards logical computation capabilities has triggered efforts on benchmarking the quality of logical qubits against physical qubits~\cite{Bluvstein2024,paetznick2024,Acharya2025,reichardt2025,Rodriguez2025,reichardt2025, rines2025demonstration}.
Establishing a meaningful metric for these benchmarks is, however, far from trivial.
To date, such benchmarks have mostly relied on proxy tasks ranging from preparing elementary entangled states to implementing subroutines of large-scale quantum algorithms. 
Against this backdrop, neutral atom logical processors have recently reached credible logical capabilities, delivering substantial contributions to scientific breakthroughs in the implementations of such early fault-tolerant building blocks~\cite{Bluvstein2024, Rodriguez2025, reichardt2025, rines2025demonstration, bluvstein2026fault}. 
The demonstrated level of performance on both gate fidelities and qubits number~\cite{evered2026, Manetsch2025} shows that neutral atoms offer a promising approach for large-scale fault-tolerant quantum computing. With zonal separation ensuring different qubit roles via reconfigurable connectivity~\cite{Beugnon2007, Bluvstein2022}, architectural roadmaps are already investigating at-scale implementation of such features~\cite{Zhou2025,cain2026}. 

Even though investigating such fundamental operations is essential in order to assess the processors' performances at the logical level, they do not fully represent the complete computations that will be performed on quantum computers for utility scale~\cite{preskill2025beyond,eisert2025mind, 2025OliviaPasqal}. 
First, certain types of operations required for universal computing (such as non-Clifford operations) are notoriously known to be hard to implement at the logical level~\cite{beverland2020lower}, while they might not appear on small-scale algorithmic demonstrations~\cite{Quantinuum2026Yamamoto}, or the preparation of elementary entangled states. 
Second, the quantum processor is expected to represent a resource in a wider hybrid quantum-classical algorithm~\cite{seelam2026qc-hpc}, and hence the quality of the application-level metrics may respond differently than pure circuit fidelity. 
While not yet achieving practical utility, the execution of elementary end-to-end quantum algorithms thus serves a twofold purpose: quantitative characterization of device behaviour in different realistic regimes, and detection of operational constraints.
In order to perform an \emph{end-to-end} comparison between logical and physical computations, three key properties are required: 
(i) sensitivity to the errors targeted by the chosen logical encoding, 
(ii) applications that are industrially meaningful and
(iii) a clear identification of what metric truly impacts the quality of the results for the chosen application.

\begin{figure}[t]
\includegraphics[width=85mm]{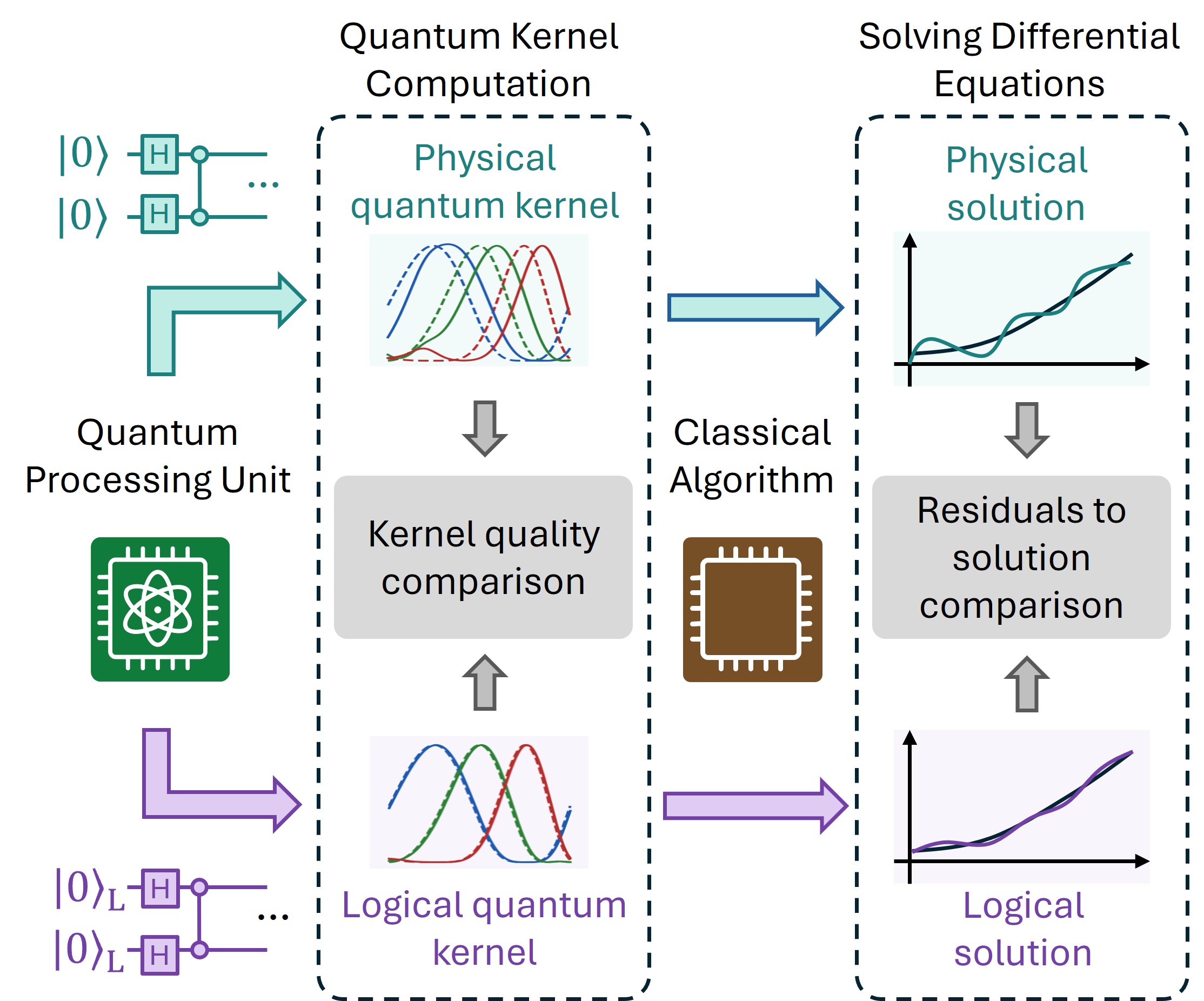}
\caption{\textbf{End-to-end benchmark comparing physical and logical quantum computations.} 
We adopt a quantum kernel as the algorithmic routine to implement using either physical qubits or logical qubits. The performances upon implementation are compared both at an intrinsic level, in terms of attained kernel quality, as well as at an applicative level, when using the quantum kernel output to solve sets of differential equations within a hybrid quantum-classical framework.}
\label{fig:fig1}
\end{figure}

In this work, and to our knowledge for the first time in the context of quantum machine learning, 
we compare logical to physical qubit implementations in a benchmark fulfilling all the three properties above, by computing quantum kernels (QKs)~\cite{Havlek2019, Schuld2019, 2021Kusumoto, Schuld2021, yin2025experimental} to solve differential equations (DEs)~\cite{Paine2023}. 
The quality of QKs, constructed via state overlaps, is by nature sensitive to error channels, such as distortions induced by coherent errors~\cite{Heyraud2022Noisy}.
Further, solving DEs provides a realistic task typically defined on continuous domains~\cite{braun1983differential}, which is highly sensitive to kernel quality, making it a robust test of kernel accuracy and an ideal setting to highlight the benefits of logical-level computation on an industrially-relevant task.
We compare the physical and logical performances both at an algorithmic primitive and application level, as summarised in Figure~\ref{fig:fig1}, using a neutral atom logical processor. 
We obtain that logical computations systematically outperform the physical ones, especially at the application level. Crucially, we observe that the logical encoding absorbs the impact of coherent errors that are not compensated by our hybrid protocol, leading to an improvement of performances that eludes the expectations based upon circuit fidelity alone. 

The structure of this work is as follows.
We first provide a motivating description of quantum kernels and how they can serve the purpose of solving differential equations.  
We then provide an overview on the architecture of our logical quantum processor, and how it can be leveraged to compute both physical and logical QKs. 
We conclude by performing a quality comparison of these kernels upon their application to the end-task of solving DEs.

\section{Solving differential equations using quantum kernels\label{sec:kernel}}

Solving DEs is a crucial task in scientific and engineering disciplines due to their ability to model natural phenomena, with a breadth of applications ranging from mechanics to biology~\cite{braun1983differential, goodwine2010engineering}. 
While classical numerical methods for solving DEs is a mature field deploying powerful techniques like finite element methods and spectral solvers, 
challenge is still posed by factors like high dimensionality, nonlinearity and multi-scale~\cite{Gear1981,Knio2006,Dolgov2020}. 
Quantum algorithms have been proposed to address such shortcomings: early proposals exclusively targeted fault-tolerant architectures via linear-algebra approaches~\cite{2009Harrow, Berry2014jeh} and guaranteed quantum speedup for specific problems, at the cost of high resource requirements. More recent proposals have also explored variational quantum machine learning approaches~\cite{Schuld2019, lubasch2020variational, 2021Oleksandr, Paine2023}, which are more amenable to implementation on near-term quantum hardware~\cite{2024Pool, 2025Schillo, philip2025experimental, 2024Jaffali}, though they only provide heuristic performance guarantees. 

We employ a quantum machine learning approach~\cite{Paine2023} inspired by classical works adopting kernels to solve differential equations~\cite{stepaniants2023learning, jalalian2025data, batlle2025error}. 
A kernel $\kappa(x,a)$ is a mathematical object which quantifies similarity between two data points $x$ and $a$, with a wide range of applications in machine learning~\cite{shawe2004kernel, hofmann2008kernel, mehrkanoon2012approximate,marquardt1994computational, scholkopf2004kernel} as we further detail in Appendix~\ref{app:kernels}.
Here, the kernel $\kappa(x,a)$ is defined on 1D variables ($x,a \in \mathbb{R}$) and used as a basis to approximate the target solution $f_{\text{sol}}(x)$ of a differential operator $\mathscr{D}(x, f(x), f'(x), \ldots, f^{(n)}(x))$. Formally, we employ mixed-model regression where $f_{\text{sol}}(x)$ is approximated by a trial solution $f_{\text{tr}}(x)$ as a linear combination of $\set{\kappa(x, a_i)}$: 
\begin{equation}
    f_{\text{tr}}(x; w_1, w_2, w_3, c) = c + \sum_{i=1}^3 w_i \kappa(x,a_i),
    \label{eq:mmr}
\end{equation}
where we use $\set{a_i} = \set{0.25,0.5,0.75}$ as the kernel centres and $w_1, w_2, w_3, c \in \mathbb{R}$ are tunable classical parameters. 
These parameters are trained to minimise a loss function $\mathscr{L}$ constructed using the differential equation, representing the discrepancy between $f_{\text{sol}}$ and $f_{\text{tr}}$.
The approach we follow is illustrated in Figure~\ref{fig:fig2}b, with details in Appendix~\ref{app:de_benchmark}.

Kernel methods can be extended employing quantum toolsets~\cite{Schuld2021}: an approach that has seen many experimental implementations~\cite{Havlek2019, 2021Kusumoto, yin2025experimental} and is expected to outperform its classical counterparts in selected instances~\cite{Havlek2019, Liu2021, Gil-Fuster2024}.
In this work, for the first time to our knowledge we use experimentally such a \emph{quantum} kernel to solve DEs. 
We define here the quantum kernel as $\kappa(x, a) \coloneq |\braket{\psi(a)|\psi(x)}|^2$, which satisfies the properties of a kernel~\cite{Schuld2021} and is computed by implementing the circuit shown in Figure~\ref{fig:fig2}a on our processor, described in the next section. 
The quantum kernel circuit is essentially composed of two unitaries: (i) a first unitary $U(x)$ with $x$ encoded into the angle of gates, and (ii) the inverse unitary $U^\dagger(a)$ with $a$ encoded the same way. Here we choose $U$ such that $f_\text{sol}$ can be correctly expressed using the computed kernel (see Appendix~\ref{app:kernel_ux_selection} for details). The kernel is computed from the fraction $p_{00}$ of $00$ bitstrings measured at the end of the circuit, using the fact that:
\begin{multline}
    |\braket{\psi(a)|\psi(x)}|^2 = \\
    \left(\bra{0} U^\dagger(x) U(a)\right) \left(\ket{0}\bra{0}\right) \left(U^\dagger(a) U(x) \ket{0}\right).
    \label{eq:qkernel}
\end{multline}
In particular, for $x=a$, we obtain that $\kappa(x,a) = |\bra{0}U^\dagger(x)U(x)\ket{0}|^2 =1$. 
Note how, referring again to Figure~\ref{fig:fig2}b, the quantum kernel as attained above is used as a (fixed) \emph{resource} of a wider, hybrid quantum-classical algorithm, in virtue of the workflow which solves the DE via mixed-model regression, with the parameter tuning running on a classical device.

\begin{figure}[t!]
\centering
\includegraphics[width=\linewidth]{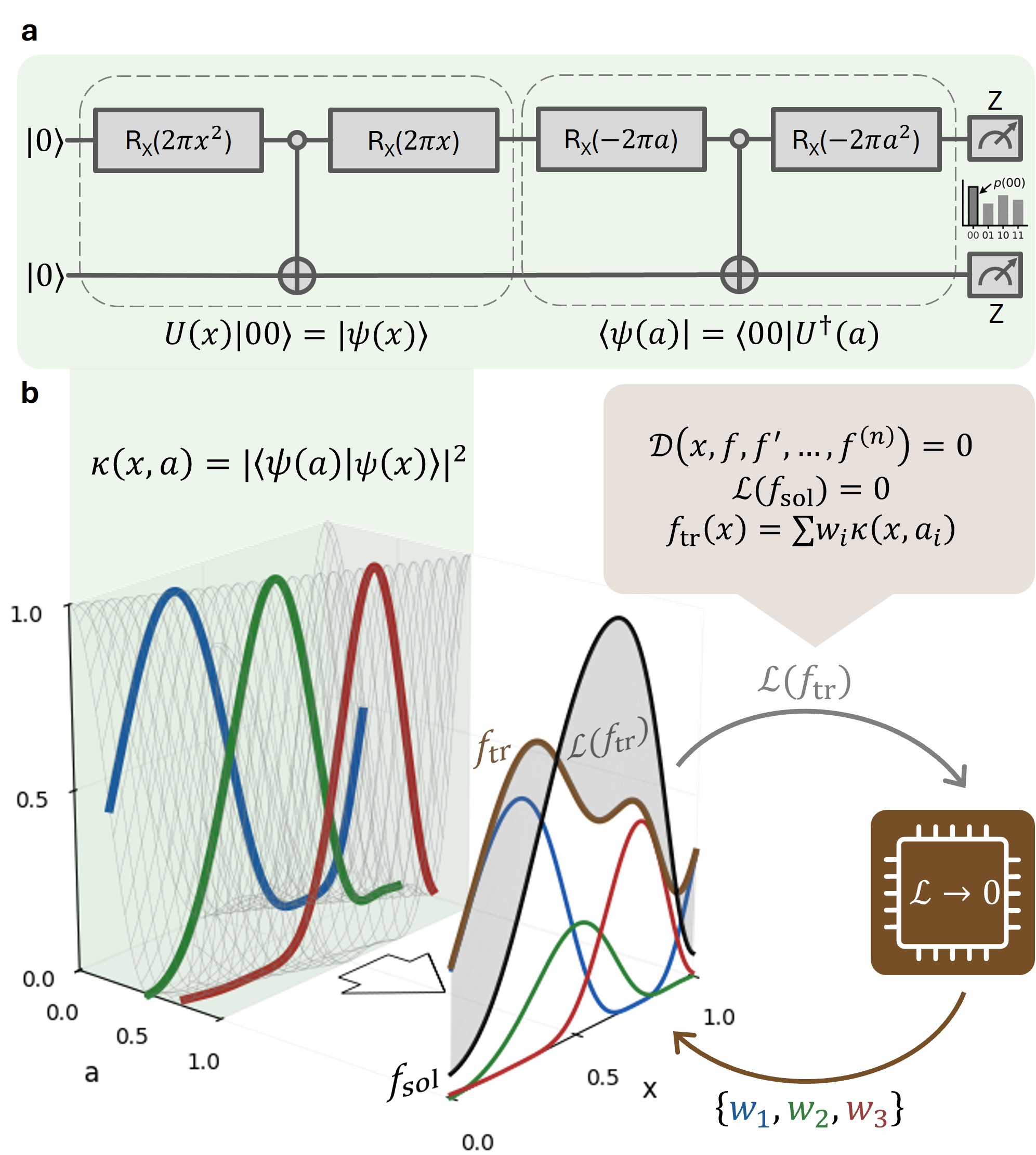}
\caption{\textbf{Solving differential equations using a quantum kernel.} 
\textbf{a.} The quantum kernel $\kappa(x,a)$ is computed by measuring the $00$ bitstring probability $p_{00}$ at the output of the displayed quantum circuit, which is equal to the inner product shown. 
\textbf{b.} Workflow of the differential equation solving based on the computed quantum kernel via mixed-model regression. The quantum kernel $\kappa(x,a)$ with three chosen parameters $\{a_i\}$ (blue, green and red curve) is used as a basis to generate a trial function $f_{\text{tr}}$.
A loss function $\mathscr{L}$ evaluates the distance of $f_{\text{tr}}$ to the solution $f_\text{sol}$ of the equation formally defined by the differential operator $\mathscr{D}$. $\mathscr{L}$ is then minimized using a classical algorithm to update the weights $\{w_i\}$ of the basis functions, thus approximating $f_\text{sol}$.
}
\label{fig:fig2}
\end{figure}

\begin{figure*}[t!]
\centering
\includegraphics[width = \linewidth]{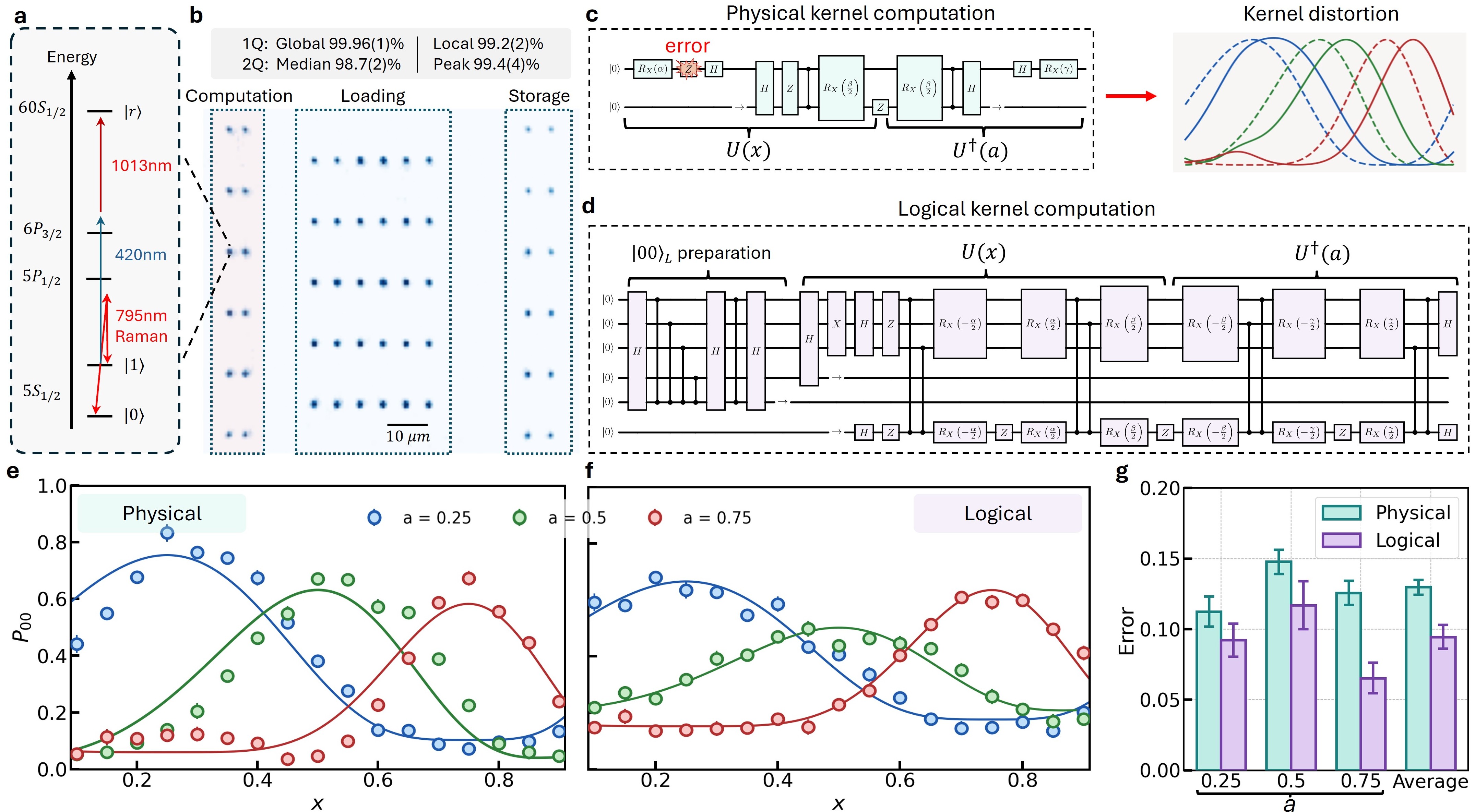}
\caption{\textbf{Quantum kernel computation using a neutral atom logical processor.} \textbf{a.} Level diagram of the relevant atomic transitions used in this work. 
\textbf{b.} Average fluorescence imaging of the atom array. We divide the space into three zones: a computation zone, a loading zone, and a storage zone. 
\textbf{c.} Circuit used to compute the physical quantum kernel. 
Distortions in the kernel shape can be traced back to coherent errors in the gates execution. 
\textbf{d.} Circuit used to compute the logical quantum kernel. It consists of three parts: preparing the logical state $\ket{00}_L = (\ket{0000} + \ket{1111})/\sqrt{2}$, performing the $U(x)$, and the $U^\dagger(a)$ operations. 
\textbf{e., f.} Experimentally attained values (markers) for the physical and logical quantum kernels $\kappa(x,a)$, compared to ideal output after rescaling and shifting (solid lines). 
\textbf{g.} Root mean square error between the rescaled and shifted ideal output versus the experimental data as in e, f. 
The error is lower at the logical level, demonstrating a closer match with the ideal output. Error bars represent the standard error on the mean and are often smaller than the symbol size.
}
\label{fig:fig3}
\end{figure*}

QKs are subject to certain roadblocks and limitations~\cite{
Shaydulin2022, Thanasilp2024, Wang2024concentration, sarkar2025concentration, agliardi2024mitigating, zhou2020limits, Thomas2025transition}, which are discussed in more detail in Appendix~\ref{app:quantum_kernel_limitations}. 
One such limitation stems from the fidelity of the quantum processor: overlap estimates as in Eq.~\ref{eq:qkernel} are evidently dependant on reliable state preparation and altered readouts can in turn invalidate the successful application of the kernel to the intended task. We distinguish two types of kernel alteration: (i) a rescaling and/or shifting of its \emph{amplitude}, e.g. when $|\braket{\psi(a)|\psi(x)}|^2<1$ for $x=a$ but still remains the maximum, and (ii) less trivial alterations affecting its \emph{shape}, which we refer to as kernel distortion (see Figure~\ref{fig:fig3}c). Such distortions can be caused by coherent errors, such as under- or over-rotations induced by gates miscalibrations. For example, an error $\varepsilon$ in the encoded unitary $U(a + \varepsilon)$ instead of $U(a)$ induces that the condition $|\braket{\psi(a+ \varepsilon)|\psi(x)}|^2 = 1$ is attained for $x= a + \varepsilon$ instead of $x=a$.

By inspecting Eq.~\eqref{eq:mmr}, it can be seen how the model for $f_{\text{tr}}$ completely absorbs into the learned classical parameters alterations of the first kind, by rescaling (via $\set{w_i}$) and shifting (via $c$) the computed kernel. 
We note that this observation is generally true beyond this work, as many hybrid quantum algorithms rely upon inexpensive classical post-processing, to absorb a certain class of alterations in a similar fashion. 
However, errors leading to shape alteration would require more complex post-processing, like error mitigation techniques, which are computationally expensive when performed at-scale~\cite{piveteau2021mitigation, zhang2025mitigation}. 
Therefore, these errors are expected to be the most harmful for our chosen workflow. Computing the kernel using logical qubits should in principle reduce such distortion, which we experimentally validate below.

\section{Computing quantum kernels on a neutral-atom logical processor}
\label{sec:hw}

We first detail the quantum kernel computations on our neutral-atom processor, based on the individual manipulation of tweezer-trapped rubidium-87 atoms. 
A similar processor was previously used to perform analog-only operations~\cite{Albrecht2023}, but it is here upgraded to support digital and logical computations, as further detailed in Appendix~\ref{app:hw}. 
For such tasks, we use the qubit basis $|0\rangle = |5S_{1/2},F=1,m_F=0\rangle$ and $|1\rangle = |5S_{1/2},F=2,m_F=0\rangle$ --- see Figure~\ref{fig:fig3}a, which exhibits in our processor a relaxation time $T_1 \simeq 5.0(2) \, \text{s}$, and a coherence time under dynamical decoupling $T_2 = 1.49(8) \, \text{s}$. 
We drive qubits via a stimulated Raman process, using a laser at 795nm which is amplitude-modulated (at the qubit transition frequency) using a fibered Mach-Zehnder interferometer~\cite{Levine2019,Levine2022}. Using this method, we reach a typical single qubit gate fidelity of $99.96(1)\%$, measured through randomized benchmarking. We perform two-qubit operations using the Rydberg blockade mechanism~\cite{Jaksch2000, Gaetan2009, Urban2009}, with the Rydberg state $|r\rangle =|60S_{1/2},m_J=1/2 \rangle$. We connect  $|1\rangle$ to $|r\rangle$ via a two-photon stimulated transition through the $6P_{3/2}$ state, using lasers at 420nm and 1013nm. We implement controlled-Z (CZ) gates using the time-optimal gate protocol~\cite{Jandura2022, Evered2023}, and obtain a space and time median fidelity of $98.7(2)\%$ measured via global randomized benchmarking~\cite{Evered2023}, with peak fidelity measured at $99.4(4)\%$.

Our processor architecture, detailed in Figure~\ref{fig:fig3}b, is based on a zoning approach~\cite{Beugnon2007,Bluvstein2022,Bluvstein2024}, where regions of space are dedicated to specific atomic operations. The processor has three zones: (i) a computation zone, in which atoms undergo single- and two-qubit gates, (ii) a loading zone, which is used at the beginning of the computation to load atoms into the optical tweezers, and (iii) a storage zone, where atoms are not impacted by operations performed in the computation zone (see Appendix~\ref{app:hw}). Atoms are shuttled between the zones using a moveable tweezer~\cite{Barredo2016}. The computation zone is globally illuminated by the Raman and Rydberg lasers, while the storage zone is sufficiently far away not to experience any significant impact from these lasers. 
The computation zone is made of pairs of traps, with intra-pair distances of typically $3 \, \mu\text{m}$, and inter-pair distances of typically $12.5 \, \mu\text{m}$. When performing a Rydberg excitation, atoms occupying a pair of traps will realize a CZ gate, while they otherwise undergo an identity operation. Local single-qubit gates are implemented in two ways: (i) via a combination of Raman laser light pulse and movement between the computation and storage zones, and (ii) via local phase gates using the moveable tweezer, with a fidelity of $99.2(2)\%$, measured via interleaved randomized benchmarking~\cite{Evered2023}.
A typical computation is then performed in the following way: after loading, cooling and initializing the atoms in the $|0\rangle$ state~\cite{Levine2019} within the loading zone, qubits are transported between the computation zone and the storage zone depending on the circuit that is implemented.

The quantum kernels are computed by implementing the circuit shown in Figure~\ref{fig:fig2}a in two variants: a physical (logical) quantum kernel is performed on a circuit with physical (logical) qubits. 
In this work, we choose to use the [[4, 2, 2]] quantum error-detecting code, which encodes $2$ logical qubits into $4$ physical qubits and can detect any single-qubit error~\cite{gottesmanQuantumFaultTolerance2016}.
Due to its dense encoding and easy manipulation, it has been the subject of numerous experimental implementations on various quantum computing platforms~\cite{bedalovFaultTolerantOperationMaterials2024, guptaEncodingMagicState2024, reichardt2025}, and is therefore a natural choice for benchmarking our logical processor capabilities. 

For both physical and logical circuits, we perform the transpilation to our native gate set and the compilation to our architecture constraints, resulting in the circuits shown in Figure~\ref{fig:fig3}c and d (further detailed in Appendix~\ref{app:kernel_circuits}). In the case of the physical kernel, we expect the finite fidelity of our gates to induce \emph{distortions} in the kernel structure: beyond reducing the contrast of the kernel, coherent errors (such as under- or over-rotations) generally alter its shape. As previously discussed, computing the kernel using an error-detecting code should in principle reduce such distortion.

The circuit used to compute the logical QK, shown in Figure~\ref{fig:fig3}d, exhibits a significantly higher complexity than the circuit in Figure~\ref{fig:fig3}c. This difference mainly comes from two aspects: the fault-tolerant preparation of $\ket{00}_L=(\ket{0000} + \ket{1111})/\sqrt{2}$ using a flag qubit, and the unitaries $U(x)$ and $U^\dagger(a)$ which are non-Clifford operations. Such operations are notoriously hard to perform on Calderbank-Shor-Steane codes such as the one used here as they do not preserve the code-space, if performed transversally. In order to limit the resource-overhead of performing such operations, we perform ancillary-assisted gates which are partially fault-tolerant, using an extra ancillary qubit (see Appendix~\ref{app:logical_qubit_implem} for details).

The experimental results of the physical (logical) QK computations are shown in Figure~\ref{fig:fig3}e(f), against the ideal outputs, adjusted to fit the contrast achieved in the respective implementation (solid lines). We directly observe on the figure that, even though the physical QK has a better contrast, its distortion with respect to the ideal output is worse.
In order to quantitatively confirm this observation, we compare the root-mean square error between ideal output and experimental data \emph{after} minimizing it through post-processing that rescales and shifts the data, as motivated in Section~\ref{sec:kernel}. The result, illustrated in Figure~\ref{fig:fig3}g, displays how the logical kernel systematically outperforms its physical counterpart in this aspect, with average median root-mean square error of $0.094$ and $0.130$ respectively. We are able to largely reproduce the experimental behaviour of both kernels using a circuit-level error model based on \emph{ab initio} simulations of the underlying atomic operations dynamics, thus tracing down the main cause of distortion to coherent errors, e.g. induced by gate imperfections, which are absorbed by the error-detecting code (see Appendix~\ref{app:circuit_simulations} for details). 

Our experimental results show that the logical encoding is able to restore the kernel shape by detecting such errors, leading to better performances at the logical level than at the physical one. 
As previously discussed, we expect that kernel distortions can impact the accuracy of solving differential equations, as we systematically investigate below.

\begin{figure}[t!]
\includegraphics[width=85mm]{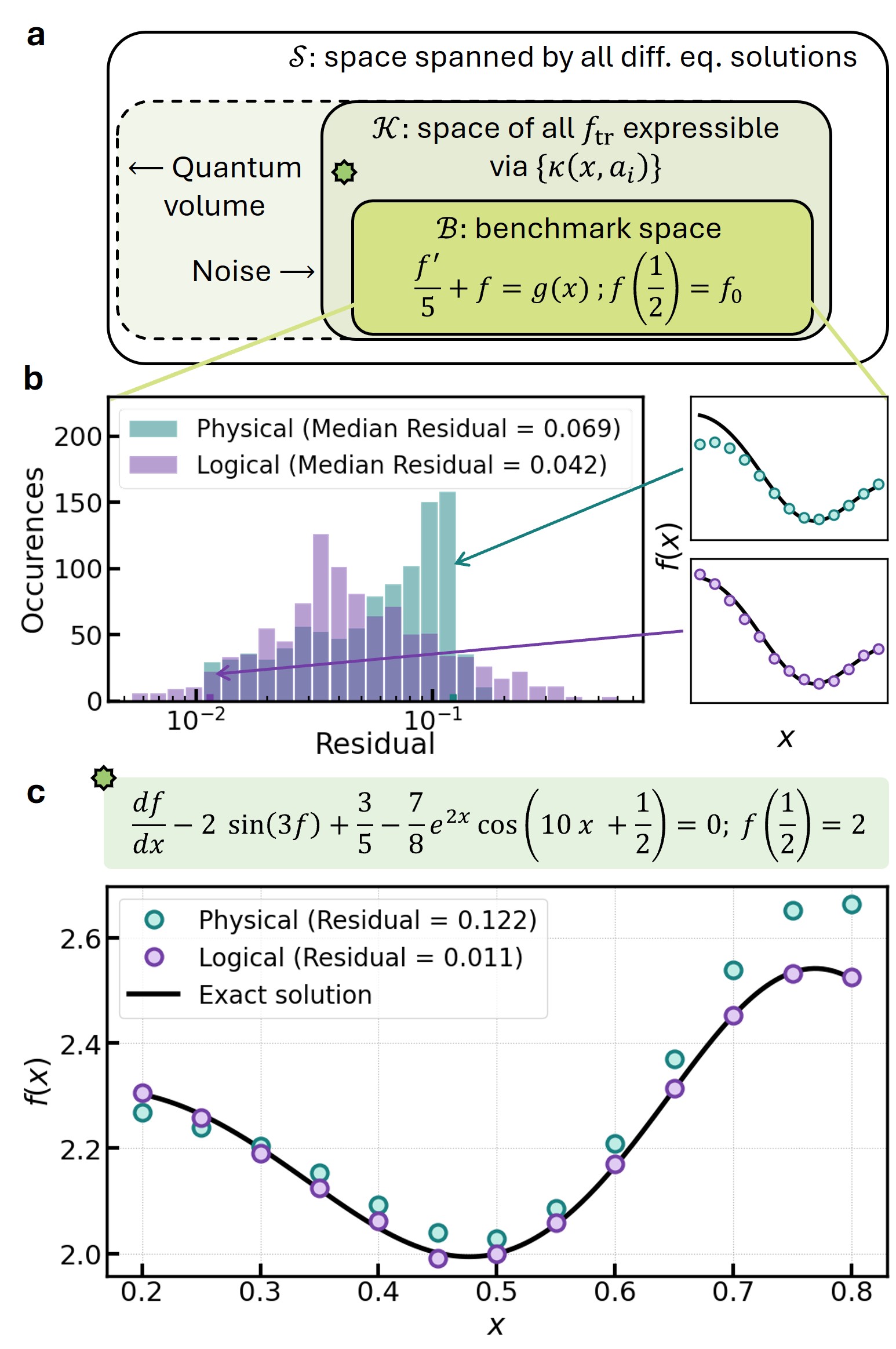}
\caption{\textbf{Solving differential equations using an experimentally-computed quantum kernel.} \textbf{a.} In the space $\mathscr{S}$ spanned by all solutions of differential equations, up to a certain accuracy quantified by some residual, there is a subset $\mathscr{K}$ that can be expressed by the kernel, which is expected to be enlarged by an increase in quantum volume and shrunk by processor error. A benchmark space $\mathscr{B}$ is defined within $\mathscr{K}$  \textbf{b.} Residual (normalized root-mean square error) while comparing the solution obtained using the kernel to the ideal solution for 1000 randomly generated differential equations from $\mathscr{B}$, along with a representative example. \textbf{c.} The kernel used to solve a nonlinear differential equation whose solution can be approximated well by the ideal kernel.}
\label{fig:fig4}
\end{figure}

\section{Solving differential equations as a kernel quality benchmark}

We now solve DEs using the obtained physical and logical quantum kernels and compare their performances. 
The method we employ to solve DEs theoretically provides an output $f_{\text{tr}}$ for any input DE. However, the \emph{residual} between $f_{\text{tr}}$ and the true solution $f_{\text{sol}}$ can remain large and $f_{\text{tr}}$ may not converge towards the true solution if $f_{\text{sol}}$ cannot be expressed by the chosen kernel $\kappa$. 
Therefore, the performance in approximating $f_{\text{sol}}$ depends crucially on the properties of the computed kernel, including the quality of its implementation. 
To quantify the discrepancy between trial and ideal solutions, we use the root-mean square error of the solution $f_{\text{tr}}$ against the exact solutions $f_{\text{sol}}$, normalized by the range of $f_{\text{sol}}$, which we henceforth refer to as the residual.

We qualitatively illustrate in Figure~\ref{fig:fig4}a how this reasoning can lead to define a \emph{space} spanned by solutions of DEs that can be tackled by our method. 
Within the space $\mathscr{S}$ spanned by the solutions of all possible DEs of a chosen dimensionality (here 1D), we define the subspace $\mathscr{K} \subset \mathscr{S}$ spanned by solutions of only those DEs which a chosen quantum kernel $\kappa$
can express up to a threshold accuracy, specified by the residual. 
When implemented on a physical device, we intuitively expect $\mathscr{K}$ to (i) expand with the quantum volume of the device, as this in general will allow the implemented kernel more degrees of freedom, and (ii) contract due to processor errors, as the kernel shape is altered. 

In order to perform a rigorous and systematic comparison between the two kernels' performances, we randomly draw differential equations from a subset $\mathscr{B} \subset \mathscr{K}$ parameterized by $\set{w_i}$, of the linear form:
\begin{align}
    \frac{1}{5}\frac{df(x)}{dx} +  f(x) - g(x, \set{w_i}) = 0, 
    \label{eq:lin_benchmark}
     \\
    f(x_0) = 2, \;\;\;  x \in [0,1],
\label{eq:lin_bcs}
\end{align}
where $g(x, \set{w_i})$ is constructed such that an ideal quantum processor with infinite shots implementing our chosen kernel would yield zero residual (see Appendix~\ref{app:de_benchmark}). We choose this subset $\mathscr{B}$ because it avoids classical trainability issues and artifacts, by ensuring a convex loss function~\cite{Paine2023}, while excluding cases where noise could inadvertently benefit certain classical or quantum learning mechanisms~\cite{1995BishopTraining,Abdolazimi2024Harnessing,Heyraud2022Noisy}. 
We solve 1000 DEs sampled from $\mathscr{B}$ using both kernels and the results are shown as a histogram in Figure~\ref{fig:fig4}b. We observe that, on average, the residual attained by the logical kernel --- which exhibits less distortion --- is more than $50\%$ lower than the one from the (distorted) physical kernel, confirming its superior performance on a use-case relevant metric. 

Though the family of first-order linear differential equation solutions in $\mathscr{B}$ provides a well-controlled space for benchmarking that is devoid of challenges arising from classical training, kernels can be applied to more challenging problems. 
To illustrate this, we include an exemplary case of a non-linear equation not admitting analytical solutions in Figure~\ref{fig:fig4}c. 
Despite its higher complexity, this example was still chosen within $\mathscr{K}$, to ensure that our ideal kernel could in principle approximate it (see Appendix~\ref{app:de_benchmark}). 
In this specific case, the solution obtained using the logical kernel exhibited a residual of $0.011$, against the physical kernel's $0.122$, a $\sim$10x difference reinforcing the previous findings.

These experimental results show that the logical kernel performs better than the physical one on metrics relevant for the task of solving differential equations. This behaviour is explained by the fact that distortions in the kernel cannot be recovered by the classical parameters of our mixed-model regression representation: we essentially incur a basis mismatch between the target space $\mathscr{K}$ and the actual space accessible to the quantum processor. 
This leads to residuals further away from the threshold theoretically expected. 
The logical kernel, which is partially protected from such distortions thanks to the logical encoding, ultimately represents a more faithful basis to decompose and express $f_{\text{sol}}$ into, leading to more reliable performances exemplified by lower residuals.

\section{Conclusion}

In this paper, we described and characterised experimentally a logical architecture for neutral atom quantum processors, and benchmarked its performance for a practical application: computing a quantum kernel and using it to solve differential equations. 

We found how our logical quantum kernel, running on a circuit equipped with error detection, outperforms its corresponding physical implementation in key metrics that quantify the quality of such kernel. 
For the first time to our knowledge, we also applied a quantum kernel to the specific task of solving differential equations in an experiment. 
Adopting standard metrics used in the physics-informed machine learning community, we observed that the superior performance of the logical quantum kernel over the physical one is confirmed also in this task. 
To support the significance of our outcomes, we ran our benchmark on a whole family of linear differential equations and an exemplary non-linear differential equation without closed-form solution.
Although this demonstration uses illustrative equations that are far from application-scale regimes --- and state-of-the-art research has so far shown prospective quantum advantage for quantum-kernel methods primarily in classification tasks --- our work addresses other immediately applicable research questions.

First, we showcase how an end-to-end logical computation for application-oriented quantum algorithms can be deployed on existing devices. 
Second, we observe how the overhead imposed by encoding logical qubits may be already compensated in current devices, when looking at metrics relevant in applicative scenarios and going beyond na\"ive comparisons of raw data, e.g. observing the contrast in the final observable attained from the quantum processor. 
Finally, the analysis of the data in controlled simulations allows us to extract and isolate the noise sources most damaging at the level of the final goal, exemplifying a case of application-informed hardware development, where such knowledge can be used to focus on how to detect or reduce such noise sources.

Going beyond this proof-of-concept work, we believe that scaling the processor gate fidelities and qubit numbers would allow us to expand the range of reachable applications of quantum kernels, and tackle more complex quantum machine-learning applications towards a regime of quantum utility for industrially-relevant tasks.

\section{Acknowledgements}
We thank the broader Pasqal team for many helpful discussions. We thank Wasiq Bokhari, Christophe Jurczak and Philippe Faist for insightful discussions. This work was supported by the PROQCIMA program within the French National Quantum Strategy (France 2030). This work was also supported by the European Union under the projects PASQuanS2.1 (HORIZON-CL4-2022 QUANTUM02-SGA, Grant Agreement 101113690), and by the Horizon Europe Programme HORIZON-CL4-2021 DIGITAL-EMERGING-01-30 via Project No. 101070144 (EuRyQa). This work was also supported by BPI under Grant Agreement DOS0254801/00 (Pasquops).

\bibliographystyle{apsrev4-2-enhanced}
\bibliography{references}

\newpage
\setcounter{figure}{0}
\renewcommand{\thefigure}{S\arabic{figure}}
\appendix

\section{Kernels and their applications\label{app:kernels}}

Kernels are conjugate-symmetric, positive (semi-) definite functions which quantify the similarity between data-points, much like inner products. Their appeal stems from the intuition that similarity measures estimated leveraging higher dimensional spaces (\emph{feature spaces}) can often bear intrinsic utility. This is illustrated in Figure~\ref{fig:appendixKernel}a for a paradigmatic task, i.e. separating two data classes via a hyperplane, which can be accomplished in the feature space upon a simple quadratic mapping, but not in the original space. 
Importantly, these similarities can be computed without explicitly carrying out the potentially expensive transformation of the input data into the feature space --- the so-called \emph{kernel trick}. 
Many problems allow a convex formulation upon application of the kernel trick, enabling the use of these methods for data and pattern analysis tasks, such as regression, classification, and correlations~\cite{shawe2004kernel}. Their practical application to real-world problems range from handwriting analysis~\cite{marquardt1994computational} to computational biology~\cite{scholkopf2004kernel}. 
When compared with other classical machine learning tools applied to approximate DE solutions, some kernel-based solvers displayed promise to better handle irregular domains, work under strict convergence guarantees, produce interpretable models and particularly solve inverse problems identifying DEs from sparse data~\cite{stepaniants2023learning, jalalian2025data, batlle2025error}.

\subsection{Quantum kernels: potential and  limitations
\label{app:quantum_kernel_limitations}}

\begin{figure}[t!]
\centering
\includegraphics[width=\linewidth]{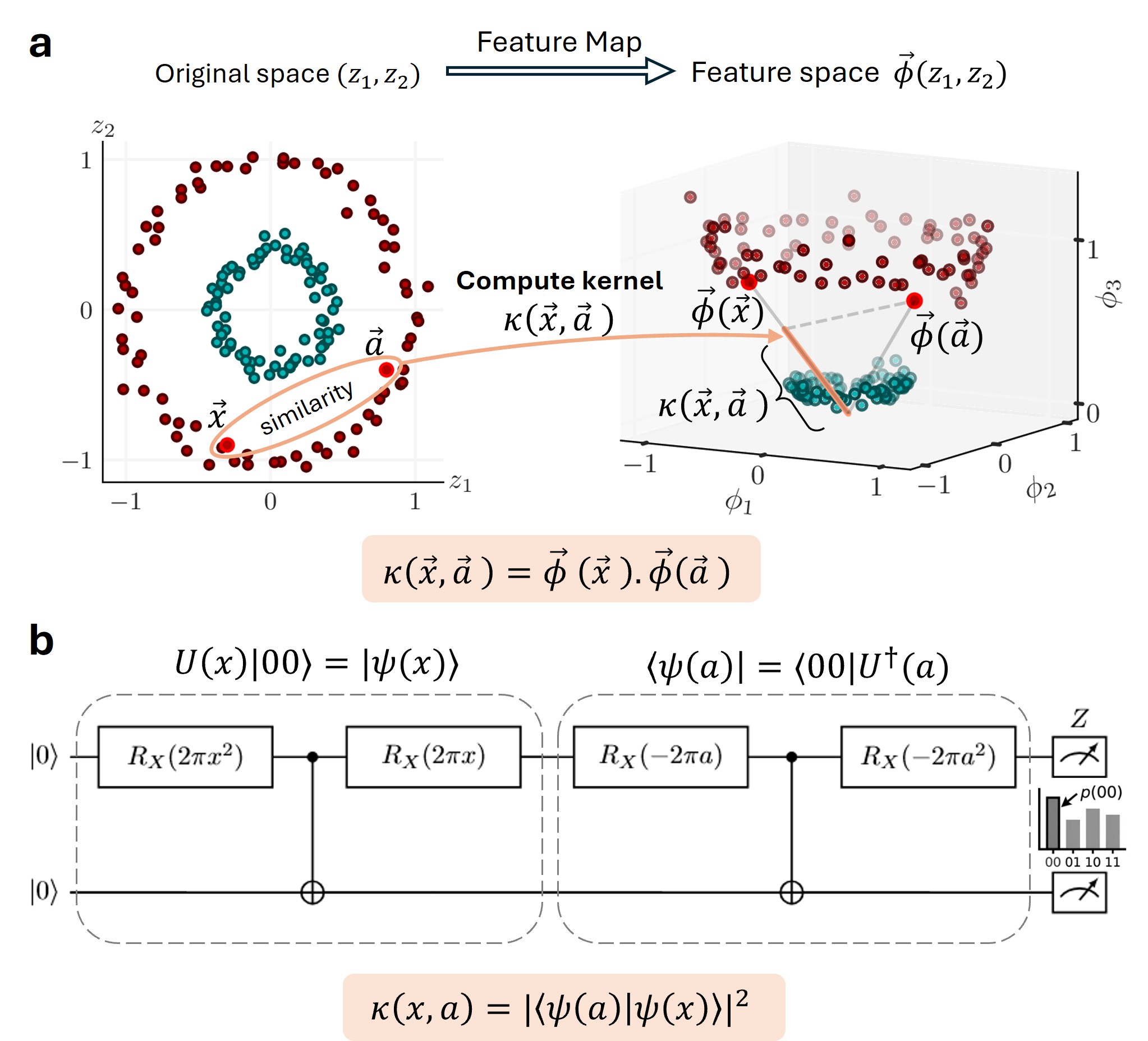}
\caption{\textbf{Parallel between classical and quantum kernel. a.} A classical kernel in a pedagogical separation example, illustrating the correspondence to the inner product in higher dimensional space. \textbf{b.} The quantum circuit in our experiment. Note that the feature space is illustrative and not the same in the two cases.}
\label{fig:appendixKernel}
\end{figure}

The large Hilbert space accessible to quantum circuits makes it a natural candidate as a feature space to map classical inputs -- and therefore to compute \emph{quantum} kernels using quantum processors~\cite{Havlek2019, Schuld2021}.
We illustrate the same example used in the main paper via Figure~\ref{fig:appendixKernel}b: though implementing a different mapping from the one in Figure~\ref{fig:appendixKernel}a, it clarifies the parallel between a kernel and a quantum kernel.

It has been demonstrated that quantum kernels can outperform their classical counterparts in selected instances~\cite{Havlek2019, Liu2021, Gil-Fuster2024}. Quantum kernels, however, also suffer from certain disadvantages. For example, recent work has highlighted that \emph{exponential concentration} can occur --- where the number of measurements required to evaluate the QK function with sufficient precision becomes prohibitively large, making them \emph{untrainable}~\cite{Shaydulin2022, Thanasilp2024}, even in presence of a reasonable amount of error mitigation~\cite{Wang2024concentration}. 
Nevertheless, this concentration does not appear for all QK functions~\cite{sarkar2025concentration} and techniques are being developed to mitigate these issues~\cite{agliardi2024mitigating}. 
One must also consider that some quantum circuits admit an efficiently computable classical surrogate: QKs described by such \emph{simulable} circuits are unlikely to prove advantageous over a purely classical kernel function~\cite{Havlek2019}.  
Also, proofs that some ideal QKs are hard to emulate classically might not hold for realistic implementations with low-fidelity circuits: approximation techniques, like tensor network methods, can frequently simulate quantum circuits under low fidelity requirements~\cite{zhou2020limits, Thomas2025transition}.

Consequently, progress towards QKs that can deliver experimental proofs of quantum advantage needs to be studied under various aspects. 
In particular, there is the necessity of finding instances that are theoretically guaranteed to be both trainable and not intrinsically simulable with classical methods~\cite{Gil-Fuster2024}. In parallel, realistic circuit fidelities must be considered, to ensure that the implemented kernel accuracy matches the requirements expected by the chosen instances. 
Fault-tolerant quantum computing being expected to provide the ultimate fidelity improvement in quantum processors~\cite{aharonov2025importance}, it can play a double role in this respect: enable the implementation of QKs intrinsically designed to run on error-corrected circuits, but also boost the performance of variational kernels theoretically implementable on near-term devices.

\subsection{The quantum kernel implemented in this work \label{app:kernel_ux_selection}}

Consider a quantum circuit expressed by a unitary $U(x)$ parametrized by the feature variable $x$, yielding the final state $\ket{\psi(x)}$. 
Then, $\kappa(x, a) \coloneq |\braket{\psi(a)|\psi(x)}|^2$ is a valid kernel function mathematically~\cite{Schuld2021}. 

When implementing a quantum kernel, the choice of the unitary $U$ is, in principle, unconstrained. We carfully choose $U$ to satisfy several design criteria: it should be compatible with hardware constraints and capabilities, admit a low-depth implementation, and avoid being separable into two independent single-qubit circuits. Moreover, over the input domain of interest, we aim to avoid a kernel landscape with multiple maxima, and we exclude constructions that are translationally invariant in $x$. Finally, we choose $U$ so that the application of logical error correction has a measurable effect on the resulting kernel.

In our implementation, $\ket{\psi(x)}$ is defined as the final state produced by the following circuit:
\begin{equation}
    \begin{quantikz}[thin lines, column sep = 0.3cm, row sep=0.4cm]
    \lstick{$\ket{0}$} & \gate{R_X(2\pi x^2)}  & \qw  & \ctrl{1} & \gate{R_X(2\pi x)} & \qw  \rstick[2]{$\ket{\psi(x)}$} \\
    \lstick{$\ket{0}$} & \qw               & \qw  & \targ{}  &  \qw               & \qw  
    \end{quantikz}.
\end{equation}

Assuming initial state $\ket{\varnothing}$ and the ability to project on $\ket{\varnothing}$ after circuit evaluation, we can circumvent the need to evaluate the overlap $\braket{\psi(a)|\psi(x)}$ by executing the deeper circuit $U^\dagger(a) U(x)$ since:  
\begin{multline}
    |\braket{\psi(a)|\psi(x)}|^2 = \left|\bra{\varnothing} U^\dagger(a) U(x) \ket{\varnothing} \right|^2,\\
    = \bra{\varnothing} U^\dagger(x) U(a) \ket{\varnothing}\bra{\varnothing} U^\dagger(a) U(x) \ket{\varnothing} .
\end{multline}
In other words, the kernel $\kappa(x, a)$ is the expectation value of the projection operator $\ket{\varnothing}\bra{\varnothing}$, acting on the state prepared as $U^\dagger(a) U(x) \ket{\varnothing}$. In our implementation we considered $\ket{\varnothing}\bra{\varnothing}=\ket{0}\bra{0}$, yielding the circuit shown in Figure~\ref{fig:fig2}a.

\subsection{Kernels for DE solving: additional comments}

When estimating trial solutions $f_{\text{tr}}$ for various data points $\{x_i\}$ in the domain, experimentally we invoke exactly the same number of shots at each $x_i$ for both the physical and logical cases (after post-selection of detected errors) to avoid shot-noise contribution biasing the comparisons.
However, we observe for example that for $\kappa$ spanning a smaller range of outputs (such as $a=0.5$), the relative effect of shot noise is higher: this limits the expectation from the infinite-shot limit, that rescaling can be completely absorbed by the classical parameter.

We remind how all the DE solutions benchmarked in the main paper are chosen to be (approximately) expressible by an ideal implementation of the kernel introduced in Eq.~\ref{eq:qkernel}.
However, we discussed how distortions in the kernel --- that cannot be recovered in the mixed-model regression state --- result in a basis mismatch between the target space $\mathscr{K}$ and the actual space accessible to the quantum processor. 
Moreover, we choose a sparse (pseudo) basis to represent the solution (via the sampled $\{a_i\}$, so it is unlikely that a basis mismatch can be treated and mitigated as a perturbation of the trial function $f_{\text{tr}}$.

At an applicative level, derivative calculations, data smoothing and noise-assisted learning can further compound the propagation of errors, and hence the interpretation of the final results. We do not investigate in detail each of these aspects involving mostly the classical training here, confining the scope to experimentally substantiate the impact of a logical implementation of quantum kernels in a complete workflow.

\subsection{Choosing the DE benchmark space and training loss functions
\label{app:de_benchmark}}

We have described in the main text the main reasoning leading to the choice of the DE family of the form in Eq.~\ref{eq:lin_benchmark}. 
First, notice how given infinite shots, and an ideal quantum processor, solutions that can be exactly expressed by our experiment need be of the form:
\begin{multline}
        h(x; \set{w_n}, c) = \\
        \sum^3_{n=1} w_n \cos^2(\pi(x - n/4)) \cos^2(\pi(x^2 - n^2/16)) + c
        \label{eq:qkernel_parameterization}
\end{multline}
for arbitrary values of $\set{w_n}$ and $c$.
To perform a fair benchmark comparing logical versus physical quantum kernels for the end-to-end task of solving a differential equation, we restricted ourselves to the space $\mathscr{B}$ spanned by the solutions described by Eq.~\ref{eq:qkernel_parameterization} and randomly sampled from it a set of equations. 
Concretely, we solved for $f(x)$ in the linear DE:
\begin{align}
    \frac{1}{5}\frac{df(x)}{dx} + f(x) &= \nonumber \\
    \frac{1}{5} \frac{dh(x)}{dx} &+ h(x) \eqqcolon g(x, \set{w_i}) \\
    f(x_0) &= h(x_0) \eqqcolon f_0 
\end{align}
where $dh(x):=dh(x; \set{w_i}, c)$ for brevity, and coefficients $w_i$ and $c$ are randomly chosen, such that $w_i \in (-1, 1) \; \forall i$ and $c\in (-1, 1)$. 
The 1/5 factor approximately balances the contribution of the two left-hand side terms in the DE, on average. 
Note that we drop the $c$ dependence of $g$ and consider $f_0$ to be arbitrary since it is not influenced by the kernel and can be absorbed by the classical training. 

We now specify the \emph{loss function} $\mathscr{L}$ chosen to capture the discrepancy between $f_{\text{sol}}$ and $f_{\text{tr}}$ in the target domain:
\begin{multline}
    \mathscr{L}(\set{w_i}, c)=\\
    \sum_{j=1}^{\Omega} \frac{1}{|\Omega|}\left(\mathscr{D}\left[x_j, f_{\text{tr}}(x_j; \set{w_i}, c),\frac{df_{\text{tr}}}{dx}\right]\right)^2 \\
    + 10 (f(x_0; \set{w_i}, c) - f_0)^2,
    \label{eq:linbench_loss}
\end{multline}
where $\set{x_j}$ is a predetermined set chosen from the domain of solution $(0,1)$ and $f_0$ is the boundary condition specified at $x_0$.
Also, note that for brevity: $\frac{df_{\text{tr}}}{dx}(x_j; \set{w_i}, c)$ is shortened as $\frac{df_{\text{tr}}}{dx}$, and $\mathscr{D}$ denotes the expression in Eq.~\ref{eq:lin_benchmark}, equated to zero, such that we expect $\mathscr{D}(\cdot)\sim 0$ for $f_{\text{tr}} \sim f_{\text{sol}}$.

\begin{figure*}[!hptb]
\centering
\includegraphics[width=\textwidth]{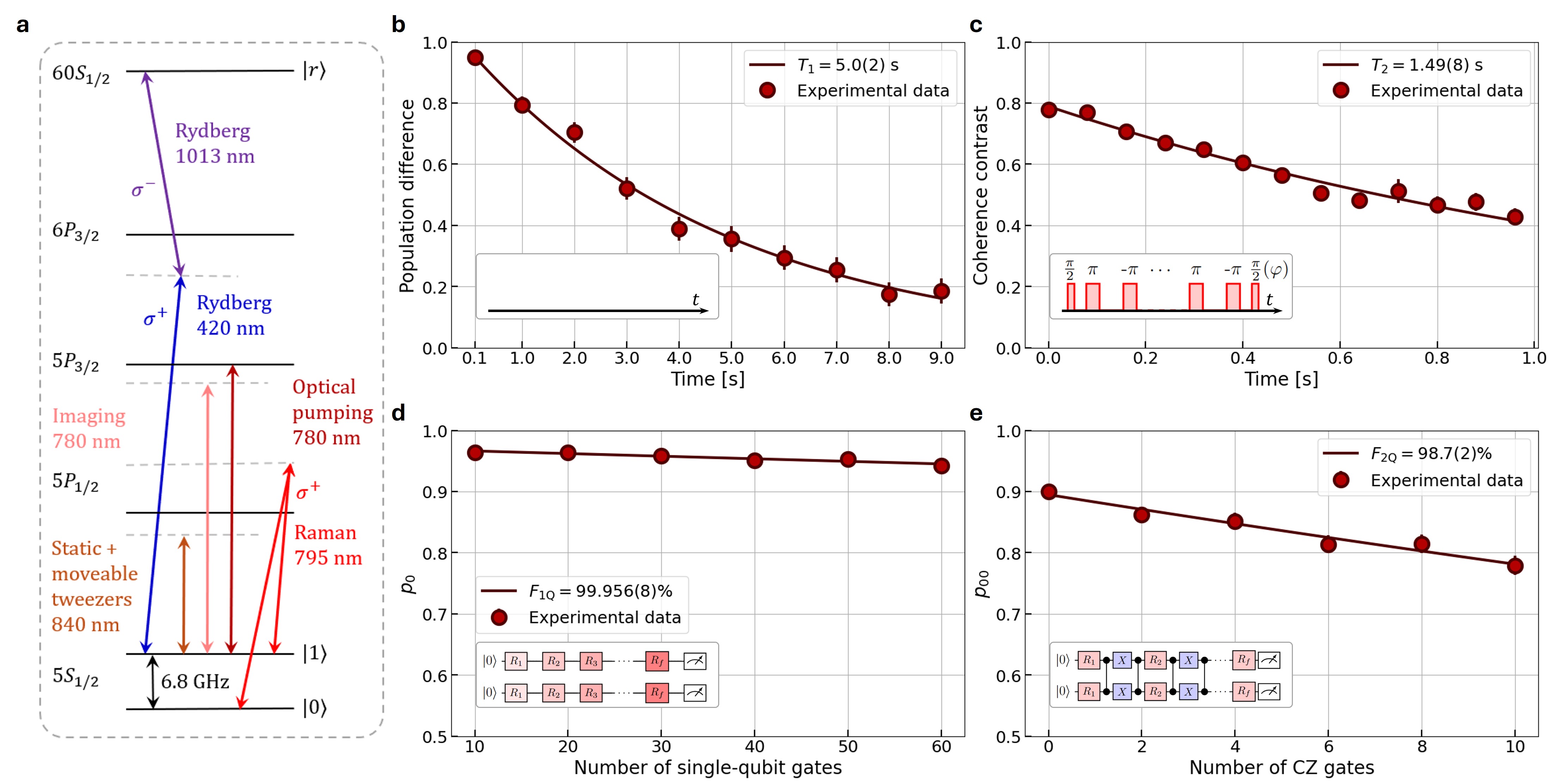}
\caption{\textbf{Precise control of rubidium hyperfine qubits. a} Atomic levels of $\text{Rb}^{87}$ used to control the hyperfine qubit. \textbf{b.} Relaxation time $T_1$, measured by computing the population difference in the $F=2$ level after and increasing amount of idling time, between experiments starting in $|F=1, m_F=0\rangle$ and $|F=2, m_F=0\rangle$. \textbf{c.} Effective coherence time $T_2$ under dynamical decoupling. After putting the atoms in superposition with an initial $\pi/2$ pulse, alternating $\pi$ and$-\pi$ pulses separated by $2\,\text{ms}$ are applied to cancel reversible dephasing. The phase of the last $\pi/2$ pulse is scanned to obtain a Ramsey oscillation, which gives the coherence contrast, the decay of which yields the coherence time. \textbf{d.} Randomized benchmarking of the single-qubit gates. \textbf{e.} Randomized benchmarking of the CZ gate. Random single-qubit rotations are interleaved with $\text{CZ}-X-\text{CZ}$ sequences in order to suppress the single-qubit phase that is accumulated during the CZ gate. The number of single-qubit rotations is kept constant so that the decay rate of the recapture probability is not affected by single-qubit gate errors, which, with the state preparation and measurement errors, offset the curve by an amount given by the data point taken with 0 CZ gates.}
\label{fig:appendixA_1}
\end{figure*}

In order to calculate the loss in Eq.~\ref{eq:linbench_loss}, of course we need access to derivatives of the trial function.
This can be expressed in terms of the derivatives of the kernel itself as
\begin{equation}
    \frac{df_{\text{tr}}}{dx} (x; \set{w_i}) = \sum_i w_i \frac{d}{dx}\kappa(x,a_i), 
\end{equation}
where $\frac{d}{dx}\kappa(x,a_i)$ may be obtained using techniques like parameter shift rule~\cite{2018MitaraiQuantum, 2019SchuldEvaluating} or numerical differentiation. In our case, we use the \texttt{interpolate.make\_smoothing\_spline} from SciPy 1.17.1 package~\cite{2020SciPy} with \texttt{lam} value of $10^{-3}$ to first create a smoothing spline and use its \texttt{derivative()} method to obtain $\frac{d}{dx}\kappa(x,a_i)$. This smoothed data and its corresponding derivatives are used in all further analysis. 

A similar procedure is followed for the non-linear DE in $\mathscr{K}$ reported in Fig.~\ref{fig:fig4}c, after making sure that the $h(x; \set{w_i}, c)$ as in Eq.~\ref{eq:qkernel_parameterization} --- stemming from our chosen kernel --- can lead to a reasonably low value of normalized root-mean square error ($0.018$), or in other words that it is possible to train $f_{\text{tr}}\sim f_{\text{sol}}$, in an ideal scenario. 

To obtain the solution in all cases, the parameters $\set{w_i}, c$ are classically optimised to minimize this loss function (which was straightforward due to the simple nature of the problem here), as visualised in Fig.~\ref{fig:fig2}.

\section{Hardware details} \label{app:hw}

Our quantum processor is based on neutral $^{87}$Rb atoms. Atoms are pre-cooled in a Zeeman slower and captured in a three-dimensional magneto-optical trap (3D-MOT), then stochastically loaded into an optical tweezer array with a single-atom occupation probability of $\approx 50\,\%$. The traps operate at $840\,\mathrm{nm}$ with a beam waist of $\approx 1\,\mathrm{\mu m}$; the static traps are generated by a spatial light modulator and focused through a pair of aspheric lenses. Defect-free target geometries are assembled, and in-circuit transport is performed using a single moving tweezer steered by a pair of crossed acousto-optic deflectors. The qubit is encoded in the magnetically insensitive clock states $\ket{0} = \ket{F=1, m_F=0}$ and $\ket{1} = \ket{F=2, m_F=0}$ of the $5S_{1/2}$ ground manifold. The different wavelengths that are used to control the qubit are presented in Figure \ref{fig:appendixA_1}a. Qubit initialization into $\ket{0}$ follows the optical pumping protocol defined in~\cite{Levine2019}, with a preparation fidelity of $99.4(1)\,\%$ at a bias field $B = 6.2\,\mathrm{G}$ along the vertical axis. State detection is performed via destructive fluorescence readout: a resonant push-out beam ejects atoms in $\ket{F=2}$, after which the remaining $\ket{F=1}$ population is imaged with a survival probability of $99.7(2)\%$. The presence (absence) of fluorescence corresponds to $\ket{0}$ ($\ket{1}$).

\paragraph*{Qubit characterization.}
We measure a relaxation time $T_1=5.0(2)\,\mathrm{s}$  by initializing the atoms in $\ket{0}$, and computing the difference of final populations in $F=2$ between measurements starting in the $\ket{F=1, m_F = 0}$ and $\ket{F=2, m_F=0}$ states as a function of time (Figure \ref{fig:appendixA_1}b). The obtained value is consistent with a combination of (1) finite vacuum lifetime, measured to be about 30 seconds on our apparatus, and (2) the impact of Raman scattering. By performing a usual Ramsey experiment we measure a coherence time $T_2^*=3\,\mathrm{ms}$, mostly limited by the spatial light modulator flickering noise. This time was extended to $T_{2} = 1.49(8)\,\mathrm{s}$, as shown in Figure~\ref{fig:appendixA_1}c, by performing a succession of echo pulses $\pi$ and $-\pi$ (mitigating slow laser intensity fluctuations), with a period of $2\,\mathrm{ms}$ between them, which is the typical duration of one move in our quantum circuits. Because we apply one spin-echo per move during the circuits, this measurement gives a good estimation of the effective coherence time for our system.

\paragraph*{Characterization of single qubit gates.}
The fidelity of our single qubit gates is estimated through global randomized benchmarking~\cite{knill2007, xia2015}. We randomly sample 30 different sets of 60 single-qubit rotations and truncate them at n = $10$, $20$, $30$, $40$, $50$, $60$ rotations and add a final pulse $R_f$ that brings the system back to its initial state $\ket{0}$ in the absence of errors.
The typical circuit that we use is shown in the inset of Figure \ref{fig:appendixA_1}d. By fitting the data to an exponential decay, we obtain a fidelity of 99.956(8)\% (Figure \ref{fig:appendixA_1}d). This value is consistent with the 99.944(2)\% fidelity obtained via ab-initio numerical simulation including all known error sources, and is limited mainly by the beam thermal sampling and spontaneous decay from the intermediate state.

\begin{figure}[t!]
\centering
\includegraphics[width=86mm]{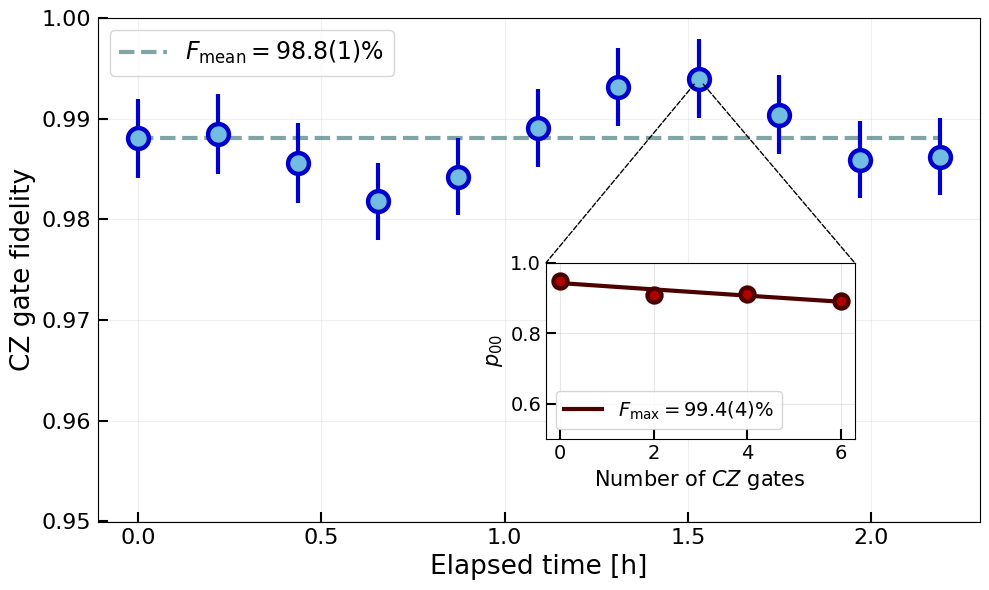}
\caption{\textbf{Two-qubit gate fidelity fluctuations.} Fluctuations of the extracted CZ gate fidelity over time for an individual qubit pair, with a peak value of 99.4(4)\%.}
\label{fig:appendix2QG}
\end{figure}

\paragraph*{Characterization of two-qubit gates.}

The two-qubit gates are implemented using the time-optimal protocol~\cite{Jandura2022, Evered2023}. The fidelity of the CZ gate is characterized using interleaved randomized benchmarking~\cite{Evered2023} where $\text{CZ}-\pi-\text{CZ}$ blocks are interleaved with random single-qubit gates. This type of randomized benchmarking is insensitive to the single-qubit phase accumulated during a CZ gate~\cite{gaebler2012, Magesan2012, Evered2023}. The exact type of gate sequence that was used is shown in the inset of Figure \ref{fig:appendixA_1}e. In a similar fashion to single-qubit gate randomized benchmarking, 30 sets of 10 single-qubit rotations are randomly sampled, and the final pulse $R_f$ is computed and added to bring the system back in the $\ket{00}$ state in the absence of errors. The number of single-qubit rotations is kept constant regardless of the number of CZ gates in the sequence, so that the decay of the return probability in the $\ket{00}$ state only depends on the CZ gate infidelity. After fitting the data with an exponential function, we obtain a fidelity of 98.7(2)\%, as shown in Figure \ref{fig:appendixA_1}e. Numerical simulations of the CZ gates yield a fidelity of 98.5\%, which is consistent with the experimental results, and show that our two-qubit gate fidelity is limited by three main factors, which are the intermediate state scattering, the shot-to-shot laser power fluctuations, and the phase noise of the lasers.

\begin{figure}[!htbp]
\centering
\includegraphics[width=85mm]{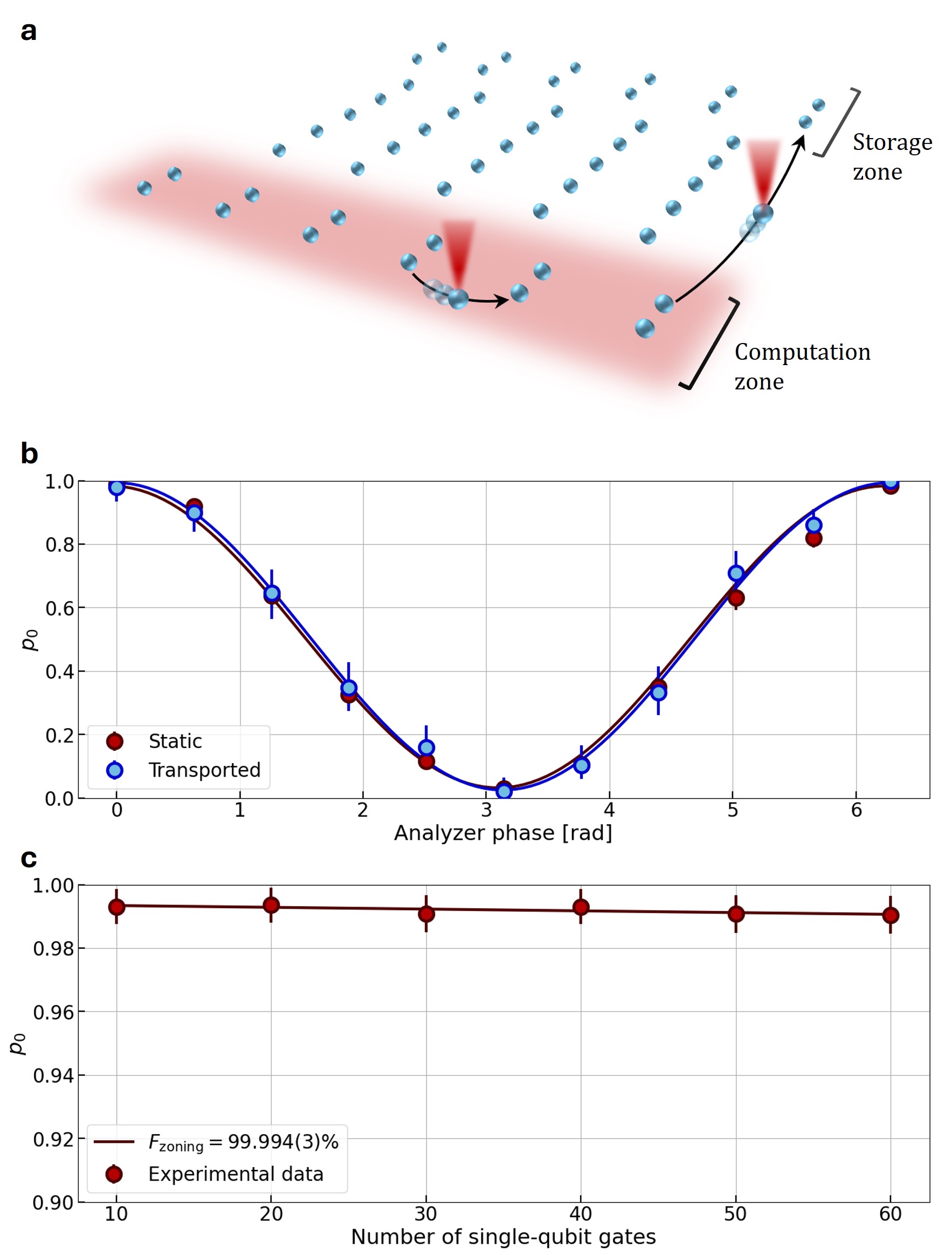}
\caption{\textbf{Register architecture and local operations. a}  Atomic layout for digital computation. The lasers that drive single- and two-qubit gates are focused on the so-called \emph{computation zone}. Operation locality is enabled by combining global illumination of the computation zone with atom transport within the computation zone and between the computation zone and the \emph{storage zone}. \textbf{b.} Coherence oscillation obtained for a static atom (red) and an atom transported on a $12.5\,\text{µm}$ distance (blue). The ratio of the contrasts is $1.00^{+0.00}_{-0.02}$ \textbf{c.} Zoning fidelity, obtained by doing single-qubit gate randomized benchmarking \emph{in the computation zone} and looking at how an atom located \emph{in the storage zone} is affected.}
\label{fig:appendixA_3}
\end{figure}

When looking at interleaved randomized benchmarking results on individual qubit pairs over an extended period of time, we observe temporal fluctuations of the extracted gate fidelity which we attribute to slow experimental drifts occurring on timescales comparable to the duration of several gate fidelity measurements, as pictured in Figure \ref{fig:appendix2QG}. The mean fidelity value for this individual qubit pair is $98.8(1)\%$ (similar values are obtained for the different individual pairs), which is compatible with the global two-qubit gate fidelity obtained over the whole computation zone. These fluctuations exhibit a peak value of $99.4(4)\%$, indicating that by improving the processor stability we could reach higher gate fidelities, which we leave for future work.

\paragraph*{In-circuit movement.} 
The ability to move atoms while fully preserving their quantum state is crucial for the implementation of quantum circuits, as atoms are shuttled to enable both arbitrary connectivity and operation locality (Figure \ref{fig:appendixA_3}a). In order to benchmark the fidelity of coherent transport inside of the computation zone, we measure the coherence of an atom over a distance of $12.5\,\text{µm}$ (typical distance between adjacent pairs) with a $\pi$-pulse applied halfway through the transport and compare it with the static case. A contrast ratio of $1.00^{+0.00}_{-0.02}$ is obtained, showing that the coherence of the state was preserved during transport. The absence of phase shift between the oscillations obtained when the atom remains static and when it is transported indicates good control over the accumulated phase during atom transport (Figure \ref{fig:appendixA_3}b).

\paragraph*{Zoning efficiency.}
In our processor architecture, we implement local single-qubit gates by combining global illumination with atom shuttling into and out of the computation zone. We characterize the efficiency of the zoning architecture by making sure that the operations performed in the computation zone have a negligible impact on atoms in the storage zone. The Raman beam has a radius waist of approximately $40\,\text{µm}$ and is centered on the computation zone, which is composed of two columns of atoms separated by $3\,\text{µm}$. The storage zone is located $51.5\,\text{µm}$ away from the computation zone. Characterization of zoning efficiency is performed by initializing an atom in the computation zone, transporting it to the storage zone, and then performing single-qubit randomized benchmarking, as described above. If the two zones are far enough from each other, the recapture probability in the $\ket{0}$ state should remain constant and close to 1 since the atom does not experience any gate. Otherwise, the atom would undergo small-angle single-qubit rotations, which do not correspond to the return pulse, and reduce the recapture in the $\ket{0}$ state. We measure a zoning efficiency of 99.994(3)\% (Figure \ref{fig:appendixA_3}c).

\section{Implementation of quantum kernels on the hardware\label{app:kernel_circuits}}
We detail how to compute a QK with circuit architecture envisaged here. The task is to implement the circuit shown in Figure~\ref{fig:fig2}a on our processor at physical and logical level using our native gate set.

\subsection{Physical circuit}
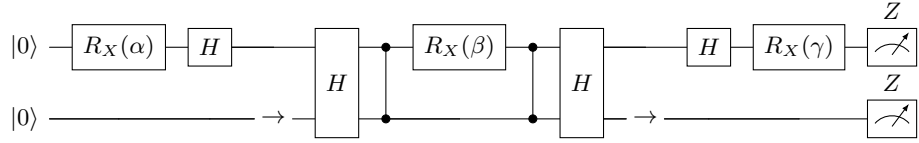
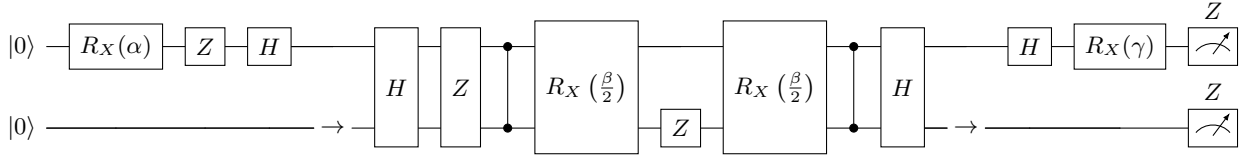
\begin{figure*}[!htbp]
    \centering

    \subfloat[The circuit in Figure~\ref{fig:fig2}a compiled for hardware execution, where $(\alpha, \beta, \gamma)=2\pi(x^2, x - a, -a^2)$. The $\rightarrow$ symbol identifies the moments where a qubit is moved in or out of the computational zone. 
    \label{fig:kernel_circuit_physical}]{
        \begin{quantikz}[thin lines, column sep = 0.3cm, row sep=0.4cm]
        \lstick{$\ket{0}$} & \gate{R_X(\alpha)} & \gate{H} & \qw       &\gate[2]{H} & \ctrl{1}    & \gate{R_X(\beta)} & \ctrl{1}    & \gate[2]{H} &\qw       &\gate{H} & \gate{R_X(\gamma)} & \meter{Z}\\
        \lstick{$\ket{0}$} & \qw              & \qw      & \arrmark  &            & \control{}  &  \qw            & \control{}  &             &\arrmark  &\qw      & \qw              & \meter{Z}
        \end{quantikz}        
    }

    \vspace{1em}
    
    \subfloat[The circuit in (b), modified to use a local $Z$ gate to perform the local parametrized $R_X(\beta)$ gate.
        \label{fig:kernel_circuit_physical_compiled}]{
        \begin{quantikz}[thin lines, column sep = 0.3cm, row sep=0.4cm]
        \lstick{$\ket{0}$} & \gate{R_X(\alpha)} &\gate{Z} &\gate{H} & \qw       &\gate[2]{H} &\gate[2]{Z}& \ctrl{1}    & \gate[2]{R_X\left(\frac{\beta}{2}\right)}&\qw      & \gate[2]{R_X\left(\frac{\beta}{2}\right)} & \ctrl{1}    & \gate[2]{H} &\qw       &\gate{H} & \gate{R_X(\gamma)} & \meter{Z}\\
        \lstick{$\ket{0}$} & \qw                &\qw      &\qw      & \arrmark  &            &           & \control{}  &                                          &\gate{Z} &                                           & \control{}  &             &\arrmark  &\qw      & \qw              & \meter{Z}
        \end{quantikz}
    }
    
    \caption{Compilation of the physical kernel circuit for computing $\kappa(x,a)$.}
    \label{fig:kernel_circuits}
\end{figure*}

Because CNOT gate is not a native operation for our processor, we decompose it into CZ and Hadamard gates. Considering moving in and out of the computation zone, it results in the circuit shown in Figure~\ref{fig:kernel_circuit_physical}, which still has a local single-qubit gate in the middle that we can not perform natively. To circumvent this, we further modify the circuit to use only global single-qubit rotations and local Z gates, see Figure~\ref{fig:kernel_circuit_physical_compiled}.
A natural question concerns the phase accumulated by qubits while being shuttled and stored in the storage zone. This does not affect qubits initialized in $\ket{0}$ or measured in the Z basis, for which the phase is irrelevant.

\subsection{Logical circuit}
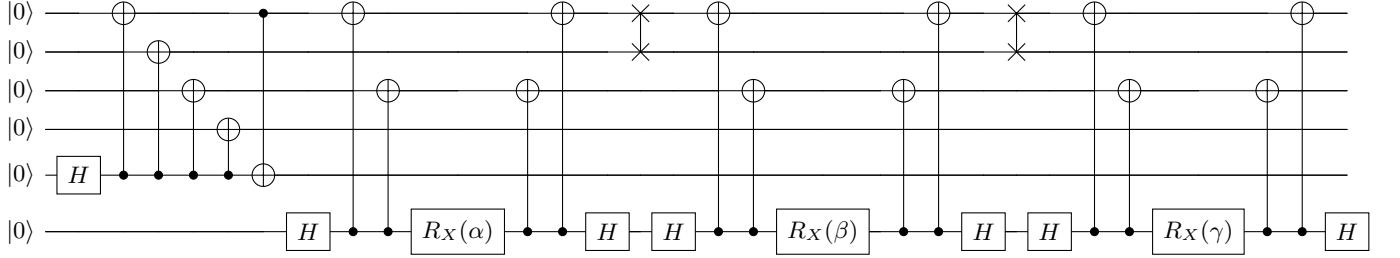
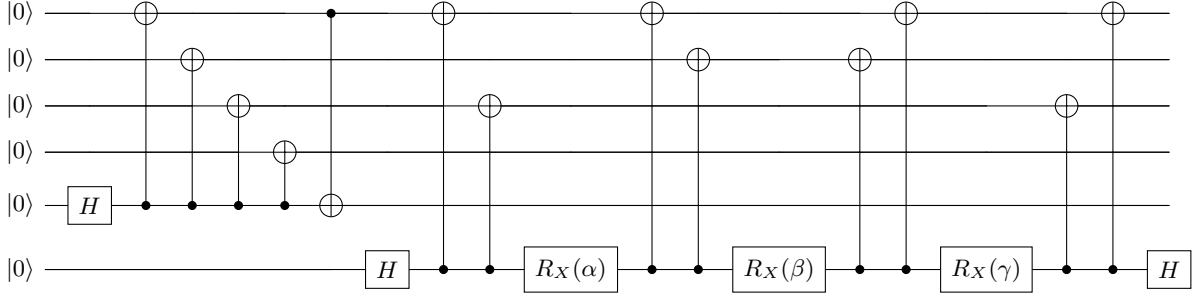
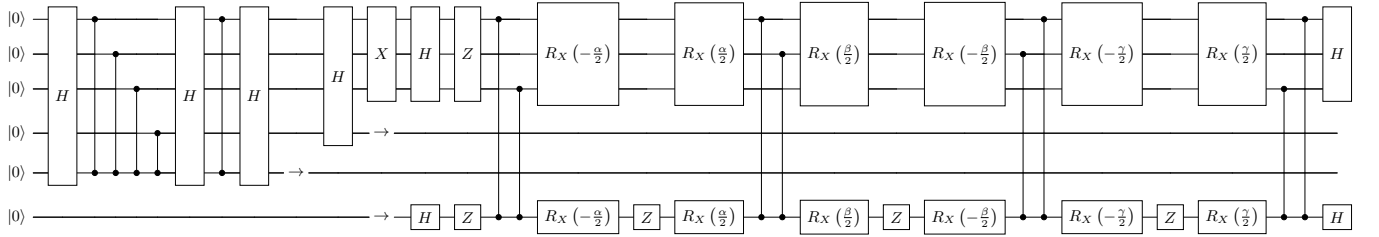
\begin{figure*}[!htbp]
    \centering
        \subfloat[The kernel circuit of Figure~\ref{fig:fig2}a encoded with the {[[4,2,2]]} code, where $(\alpha, \beta, \gamma)=2\pi(x^2, x - a, -a^2)$.
        \label{fig:kernel_circuit_encoded}]{
        \begin{quantikz}[thin lines, column sep = 0.15cm, row sep=0.2cm]
        \lstick{$\ket{0}$} & \qw      & \targ{}  & \qw      & \qw      & \qw      & \ctrl{4} & \qw      & \targ{}  & \qw      & \qw & \qw      & \targ{}  & \qw      & \swap{1} & \qw      & \targ{}  & \qw      & \qw & \qw & \qw      & \targ{}  & \qw      & \swap{1} & \qw      & \targ{}  & \qw      & \qw & \qw      & \targ{}  & \qw \\
        \lstick{$\ket{0}$} & \qw      & \qw      & \targ{}  & \qw      & \qw      & \qw      & \qw      & \qw      & \qw      & \qw & \qw      & \qw      & \qw      & \targX{} & \qw      & \qw      & \qw      & \qw & \qw & \qw      & \qw      & \qw      & \targX{} & \qw      & \qw      & \qw      & \qw & \qw      & \qw      & \qw \\
        \lstick{$\ket{0}$} & \qw      & \qw      & \qw      & \targ{}  & \qw      & \qw      & \qw      & \qw      & \targ{}  & \qw & \targ{}  & \qw      & \qw      & \qw      & \qw      & \qw      & \targ{}  & \qw & \qw & \targ{}  & \qw      & \qw      & \qw      & \qw      & \qw      & \targ{}  & \qw & \targ{}  & \qw      & \qw \\
        \lstick{$\ket{0}$} & \qw      & \qw      & \qw      & \qw      & \targ{}  & \qw      & \qw      & \qw      & \qw      & \qw & \qw      & \qw      & \qw      & \qw      & \qw      & \qw      & \qw      & \qw & \qw & \qw      & \qw      & \qw      & \qw      & \qw      & \qw      & \qw      & \qw & \qw      & \qw      & \qw \\
        \lstick{$\ket{0}$} & \gate{H} & \ctrl{-4}& \ctrl{-3}& \ctrl{-2}& \ctrl{-1}& \targ{}  & \qw      & \qw      & \qw      & \qw & \qw      & \qw      & \qw      & \qw      & \qw      & \qw      & \qw      & \qw & \qw & \qw      & \qw      & \qw      & \qw      & \qw      & \qw      & \qw      & \qw & \qw      & \qw      & \qw \\
        \lstick{$\ket{0}$} & \qw      & \qw      & \qw      & \qw      & \qw      & \qw      & \gate{H} & \ctrl{-5}& \ctrl{-3}& \gate{R_X(\alpha)} & \ctrl{-3}& \ctrl{-5}& \gate{H} & \qw      & \gate{H} & \ctrl{-5}& \ctrl{-3}& \gate{R_X(\beta)} & \qw & \ctrl{-3}& \ctrl{-5}& \gate{H} & \qw      & \gate{H} & \ctrl{-5}& \ctrl{-3}& \gate{R_X(\gamma)} & \ctrl{-3}& \ctrl{-5}& \gate{H}
        \end{quantikz}
    }

    \vspace{1em}
    
   \subfloat[The circuit in (a), optimized to reduce gate count.
        \label{fig:kernel_circuit_encoded_optimized}
    ]{
        \begin{quantikz}[thin lines, column sep = 0.3cm, row sep=0.3cm]
        \lstick{$\ket{0}$} & \qw      & \targ{}  & \qw      & \qw      & \qw      & \ctrl{4} & \qw      & \targ{}  & \qw      & \qw                & \targ{}  & \qw      & \qw               & \qw      & \targ{}  & \qw                & \qw      & \targ{}  & \qw \\
        \lstick{$\ket{0}$} & \qw      & \qw      & \targ{}  & \qw      & \qw      & \qw      & \qw      & \qw      & \qw      & \qw                & \qw      & \targ{}  & \qw               & \targ{}  & \qw      & \qw                & \qw      & \qw      & \qw \\
        \lstick{$\ket{0}$} & \qw      & \qw      & \qw      & \targ{}  & \qw      & \qw      & \qw      & \qw      & \targ{}  & \qw                & \qw      & \qw      & \qw               & \qw      & \qw      & \qw                & \targ{}  & \qw      & \qw \\
        \lstick{$\ket{0}$} & \qw      & \qw      & \qw      & \qw      & \targ{}  & \qw      & \qw      & \qw      & \qw      & \qw                & \qw      & \qw      & \qw               & \qw      & \qw      & \qw                & \qw      & \qw      & \qw \\
        \lstick{$\ket{0}$} & \gate{H} & \ctrl{-4}& \ctrl{-3}& \ctrl{-2}& \ctrl{-1}& \targ{}  & \qw      & \qw      & \qw      & \qw                & \qw      & \qw      & \qw               & \qw      & \qw      & \qw                & \qw      & \qw      & \qw \\
        \lstick{$\ket{0}$} & \qw      & \qw      & \qw      & \qw      & \qw      & \qw      & \gate{H} & \ctrl{-5}& \ctrl{-3}& \gate{R_X(\alpha)} & \ctrl{-5}& \ctrl{-4}& \gate{R_X(\beta)} & \ctrl{-4}& \ctrl{-5}& \gate{R_X(\gamma)} & \ctrl{-3}& \ctrl{-5}& \gate{H}
        \end{quantikz}
        }
    
    \vspace{1em}
    
    \subfloat[The circuit in (b), compiled for hardware execution. The $\rightarrow$ symbol identifies the moments where a qubit is moved in or out of the computational zone. All gates except for $Z$ are applied globally to all qubits in the computational zone.
        \label{fig:kernel_circuit_encoded_compiled}
    ]{\begin{adjustbox}{width=\textwidth}
        \begin{quantikz}[thin lines, column sep = 0.3cm, row sep=0.3cm, scale=0.5]
        \lstick{$\ket{0}$} & \gate[5]{H} & \ctrl{4}   & \qw        & \qw        & \qw        & \gate[5]{H} & \ctrl{4}   & \gate[5]{H} & \qw      & \gate[4]{H} & \gate[3]{X} & \gate[3]{H} & \gate[3]{Z}& \ctrl{5}   & \qw        & \gate[3]{R_X\left(-\frac{\alpha}{2}\right)}& \qw     & \gate[3]{R_X\left(\frac{\alpha}{2}\right)}& \ctrl{5}   & \qw        & \gate[3]{R_X\left(\frac{\beta}{2}\right)}& \qw     & \gate[3]{R_X\left(-\frac{\beta}{2}\right)}& \qw        & \ctrl{5}   & \gate[3]{R_X\left(-\frac{\gamma}{2}\right)}& \qw     & \gate[3]{R_X\left(\frac{\gamma}{2}\right)}& \qw        & \ctrl{5}   & \gate[3]{H} \\
        \lstick{$\ket{0}$} & \qw         & \qw        & \ctrl{3}   & \qw        & \qw        & \qw         & \qw        & \qw         & \qw      & \qw         & \qw         & \qw         &            & \qw        & \qw        &                                            & \qw     &                                            & \qw       & \ctrl{4}   &                                          & \qw     &                                           & \ctrl{4}   & \qw        &                                            & \qw     &                                           & \qw        & \qw        & \qw         \\
        \lstick{$\ket{0}$} & \qw         & \qw        & \qw        & \ctrl{2}   & \qw        & \qw         & \qw        & \qw         & \qw      & \qw         & \qw         & \qw         &            & \qw        & \ctrl{3}   &                                            & \qw     &                                            & \qw       & \qw        &                                          & \qw     &                                           & \qw        & \qw        &                                            & \qw     &                                           & \ctrl{3}   & \qw        & \qw         \\
        \lstick{$\ket{0}$} & \qw         & \qw        & \qw        & \qw        & \ctrl{1}   & \qw         & \qw        & \qw         & \qw      & \qw         & \arrmark    & \qw         & \qw        & \qw        & \qw        & \qw                                        & \qw     & \qw                                        & \qw       & \qw        & \qw                                      & \qw     & \qw                                       & \qw        & \qw        & \qw                                        & \qw     & \qw                                       & \qw        & \qw        & \qw         \\
        \lstick{$\ket{0}$} & \qw         & \control{} & \control{} & \control{} & \control{} & \qw         & \control{} & \qw         & \arrmark & \qw         & \qw         & \qw         & \qw        & \qw        & \qw        & \qw                                        & \qw     & \qw                                        & \qw       & \qw        & \qw                                      & \qw     & \qw                                       & \qw        & \qw        & \qw                                        & \qw     & \qw                                       & \qw        & \qw        & \qw         \\
        \lstick{$\ket{0}$} & \qw         & \qw        & \qw        & \qw        & \qw        & \qw         & \qw        & \qw         & \qw      & \qw         & \arrmark    & \gate{H}    & \gate{Z}   & \control{} & \control{} & \gate{R_X\left(-\frac{\alpha}{2}\right)}   & \gate{Z}& \gate{R_X\left(\frac{\alpha}{2}\right)}   & \control{} & \control{} & \gate{R_X\left(\frac{\beta}{2}\right)}   & \gate{Z}& \gate{R_X\left(-\frac{\beta}{2}\right)}   & \control{} & \control{} & \gate{R_X\left(-\frac{\gamma}{2}\right)}   & \gate{Z}& \gate{R_X\left(\frac{\gamma}{2}\right)}   & \control{} & \control{} & \gate{H}
        \end{quantikz}
        \end{adjustbox}
    }
    
    \caption{Logical encoding and compilation of the kernel circuit for calculating $\kappa(x,a)$, using the [[4,2,2]] code.}
    \label{fig:kernel_circuits_encoded}
\end{figure*}

Starting from the circuit in Figure~\ref{fig:fig2}a, we encode each part into its logical equivalent, resulting in the circuit in Figure~\ref{fig:kernel_circuit_encoded}. Notably, we start with the $\ket{00}_L$ state preparation and use proxy-phasing (see Appendix~\ref{app:logical_qubit_implem}) to perform each parametrized $R_X$ gate. Additionally, the CNOT gates in the original physical circuit are encoded as physical SWAP gates between the first and second data qubits.

At this stage, the circuit is neither simplified nor compiled for hardware execution. We first take advantage of quantum gate identities and commutation relations to eliminate as many redundant gates as possible (Figure~\ref{fig:kernel_circuit_encoded_optimized}). 

We then compile the circuit to the device's native set of operations --- similarly the compilation of the physical QK circuit --- and obtain the circuit in Figure~\ref{fig:kernel_circuit_encoded_compiled}.

\begin{figure}[!htbp]
\includegraphics[width=90mm]{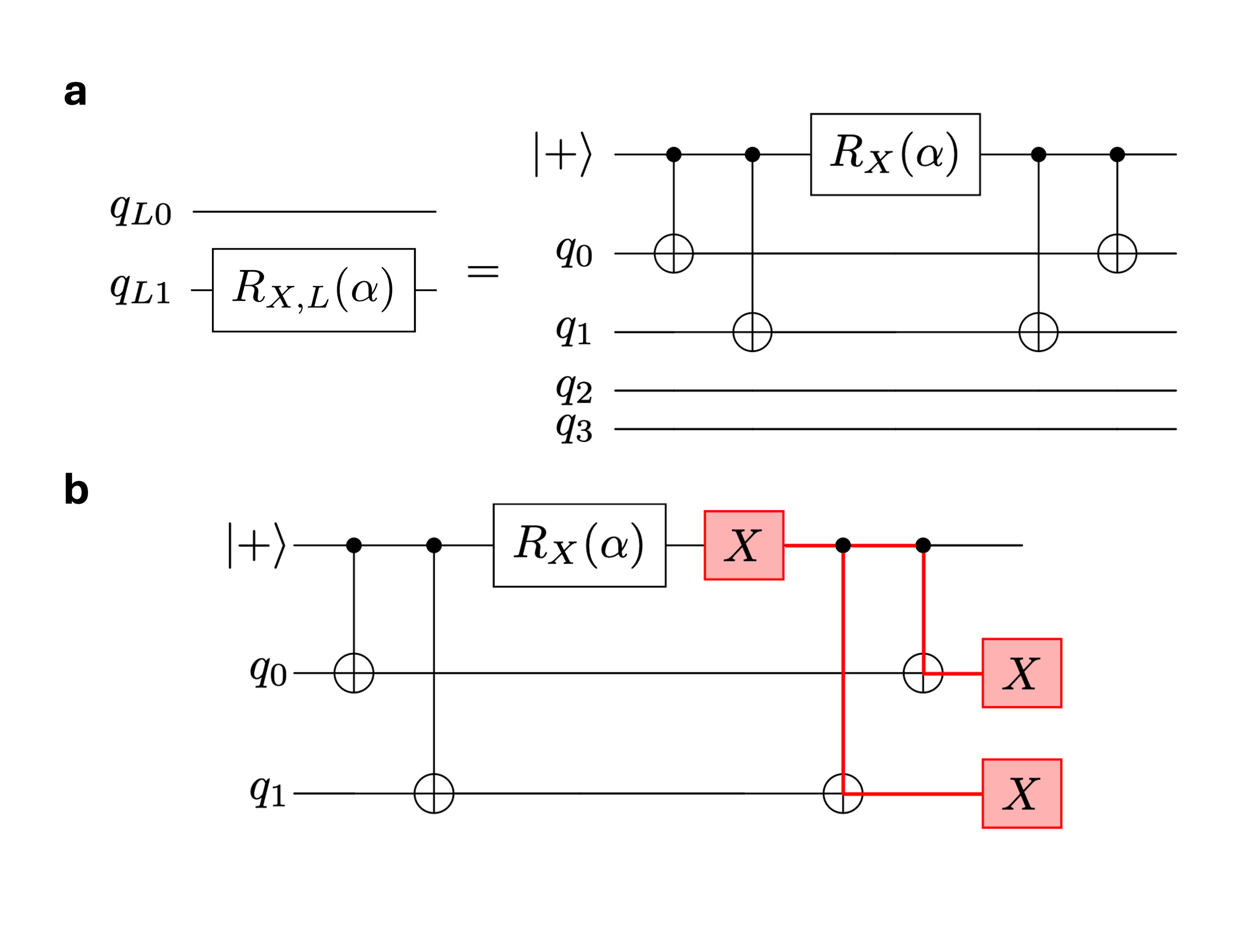}
\caption{\textbf{Logical rotation using proxy phasing.} \textbf{a} The logical rotation (left) can be implemented directly on the physical qubits or using an ancilla qubit (right). In the ancilla-based approach, a $\ket{+}$ state is prepared, entangled with the data qubits, rotated by $R_X(\alpha)$, and then disentangled. \textbf{b}Example of the non fault-tolerance of the proxy-phasing.}

\label{fig:proxy_phasing}
\end{figure}

\section{Logical qubits implementation \label{app:logical_qubit_implem}}

\paragraph*{Logical qubits.}
The [[4, 2, 2]] code is a Calderbank-Shor-Steane code defined by two stabilizer generators: $S_X = X_0 X_1 X_2 X_3 \quad \text{and}\quad S_Z = Z_0 Z_1 Z_2 Z_3$.
The logical states are given by :
\begin{align}
    \ket{00}_L &=(\ket{0000} + \ket{1111})/\sqrt{2} \\
    \ket{01}_L &=(\ket{0011} + \ket{1100})/\sqrt{2} \\
    \ket{10}_L &= (\ket{0101} + \ket{1010})/\sqrt{2} \\
    \ket{11}_L &= (\ket{0110} + \ket{1001})/\sqrt{2}
\end{align}

The fault-tolerant preparation of $\ket{00}_L$ uses a flag qubit~\cite{gottesmanQuantumFaultTolerance2016, bedalovFaultTolerantOperationMaterials2024, reichardt2025} to detect intermediate parity violations: measurement in $\ket{1}$ signals an error, prompting discard of the state and ensuring any single physical error remain detectable.

\paragraph*{Ancilla-assisted method for non-Clifford operations.} The transversal gate set of the $[[4, 2, 2]]$ code does not include the arbitrary single-qubit rotations necessary for Quantum Kernel  applications~\cite{eastinRestrictionsTransversalEncoded2009}.

To circumvent this, we employ proxy phasing~\cite{ProxyPhasingComputed} as shown in Figure~\ref{fig:proxy_phasing}, enabling arbitrary logical rotation by coupling the logical state to an ancilla qubit and applying the arbitrary rotation to the latter. Formally, let $U_{pred}$ be a unitary that computes a predicate function $f(x)$ of the state $\ket{x}$ into an ancilla initialized to $\ket{0}$:
\begin{equation}
 U_{pred} (\ket{x} \otimes \ket{0}) = \ket{x} \otimes \ket{f(x)}
\end{equation}

\noindent Applying $Z(\theta)$ to the ancilla yields:  
\begin{equation}
 \ket{x} \otimes \ket{f(x)} \xrightarrow{I \otimes Z(\theta)} e^{i \theta \cdot f(x)}(\ket{x} \otimes \ket{f(x)})
\end{equation}
Finally, uncomputing the predicate restores the ancilla to $\ket{0}$ while leaving the phase on the data state:  
\begin{equation} U_{pred}^\dagger (e^{i \theta \cdot f(x)}\ket{x} \otimes  \ket{f(x)}) = e^{i \theta \cdot f(x)} \ket{x} \otimes \ket{0}
\end{equation}

This generalises to the $X$ basis and we can implement $X_iX_j(\theta)$ gate using the joint $R_X$-parity of two qubits as a predicate function. In the context of the $[[4, 2, 2]]$ code, we can thus perform a logical $X_{1L} = X_L \otimes I_L$ rotation on the first logical qubit by computing the parity of the physical qubits that constitute that logical operator.
Given the logical operator $X_{L1}=X_0 I_1 X_2 I_3$, an arbitrary logical rotation $X_{L1}(\theta)$ is performed by employing an ancilla initialized in $\ket{+}$ and coupled via CNOTs to physical qubits 1 and 2.   

\paragraph*{Fault-tolerance limitations.} Although this provides a universal gate set, the proxy-phasing implementation used here is not fault-tolerant. If an error occurs on the ancilla qubit \emph{after} the first set of entangling CNOTs but \emph{before} the uncomputation step, this error can propagate back to the data qubits during the uncomputation CNOTs (Figure \ref{fig:proxy_phasing}b).

\section{Error modelling of the\\ quantum kernel circuits}\label{app:circuit_simulations}

To gain insight into experimental errors and their impact on the shape of the quantum kernel, we perform an \textit{ab-initio} simulation of the physical and logical kernel circuits, which takes into account the finite fidelities of all hardware operations described previously. We first detail the numerical method we adopt to simulate the circuits, then compare our error model results against our experimental results and observe an agreement between them, showing that we have a good understanding of our platform. We finish by providing an example of coherent error (induced e.g. by under- or over- gate rotation) which induces a distortion of the kernel at the physical level, which is restored at the logical level thanks to the logical encoding. 

\subsection{Details on error modelling}

The simulation takes into account all the operations performed on the hardware, including dynamical decoupling pulses and coherent moves. Each of these operations is implemented as a three-step process: ideal unitary (identity for coherent move), Pauli error channels acting on the qubit subspace, and an additional atom loss channel that takes the atom out of the qubit subspace. The simulations are performed using the Cirq quantum computing framework~\cite{Cirq_Developers_2025}. 
In order to take into account atom loss, the simulation computes the dynamics on three levels $\{|0\rangle, |1\rangle, |L\rangle\}$, where $\{|0\rangle, |1\rangle \}$ are the computational states and $|L\rangle$ represents a loss state. At the end of the dynamics, to faithfully simulate readout, we combine the $|1\rangle$ and $|L\rangle$ outcomes as both correspond to the absence of an atom during fluorescence imaging. We note that in this error model, we interpret leakage into $F=2$ (outside $\ket{1}$) as being an atom loss error, since our detection method induces the loss of these atoms. Therefore, we do not take into account leakage errors, as leakage to $F=1$ (outside $\ket{0}$) is negligible as compared to other error channels. We provide below some details on the way we extract the Pauli error channels and the atom loss error channel from \textit{ab-initio} simulations of the atomic operations.

\paragraph*{Pauli error model.} To extract the effective Pauli error channel on the computational subspace, we perform an \textit{ab-initio} simulation of the considered atomic operation, including all known noise sources coming from the processor, on a set of input density matrices spanning the computational subspace, then trace out all non-computational states from both input and output density matrices. From these reduced input-output pairs, we compute the effective quantum channel $\Lambda$ via its quantum process tomography $\chi$ matrix:

\begin{equation}
\Lambda(\rho) = \sum_{m,n} \chi_{mn} P_m \rho P_n^\dagger,
\end{equation}
where $\rho$ is the density matrix and $P_i$ is an element of $\{I, X, Y, Z\}$ for one-qubit gates or $\{I \otimes I, I \otimes X, \ldots, Z \otimes Z\}$ for two-qubit gate. By factoring out the ideal gate from $\Lambda$, we can extract the corresponding noise channel $\mathcal{E}$:

\begin{equation}
    \mathcal{E}(\rho)=\sum_i K_i \rho K_i^{\dagger},
\end{equation}

\noindent where $K_i$ is Kraus operator. This expression, after Pauli twirling approximation~\cite{emersonSymmetrizedCharacterizationNoisy2007}, becomes : 

\begin{equation}
\mathcal{E}_{\text {Pauli}}(\rho)=\sum_i p_i {P}_i \rho{P}_i
,
\end{equation}
where $p_i$ is the probability associated to the corresponding $P_i$.

\paragraph*{Atom loss.} To extract the probability of atom loss for each atomic operation, we look at the population in the states outside of the computational space, both using simulations and by performing experimental measurements.

\begin{figure}[t!]
\centering
\includegraphics[width=86mm]{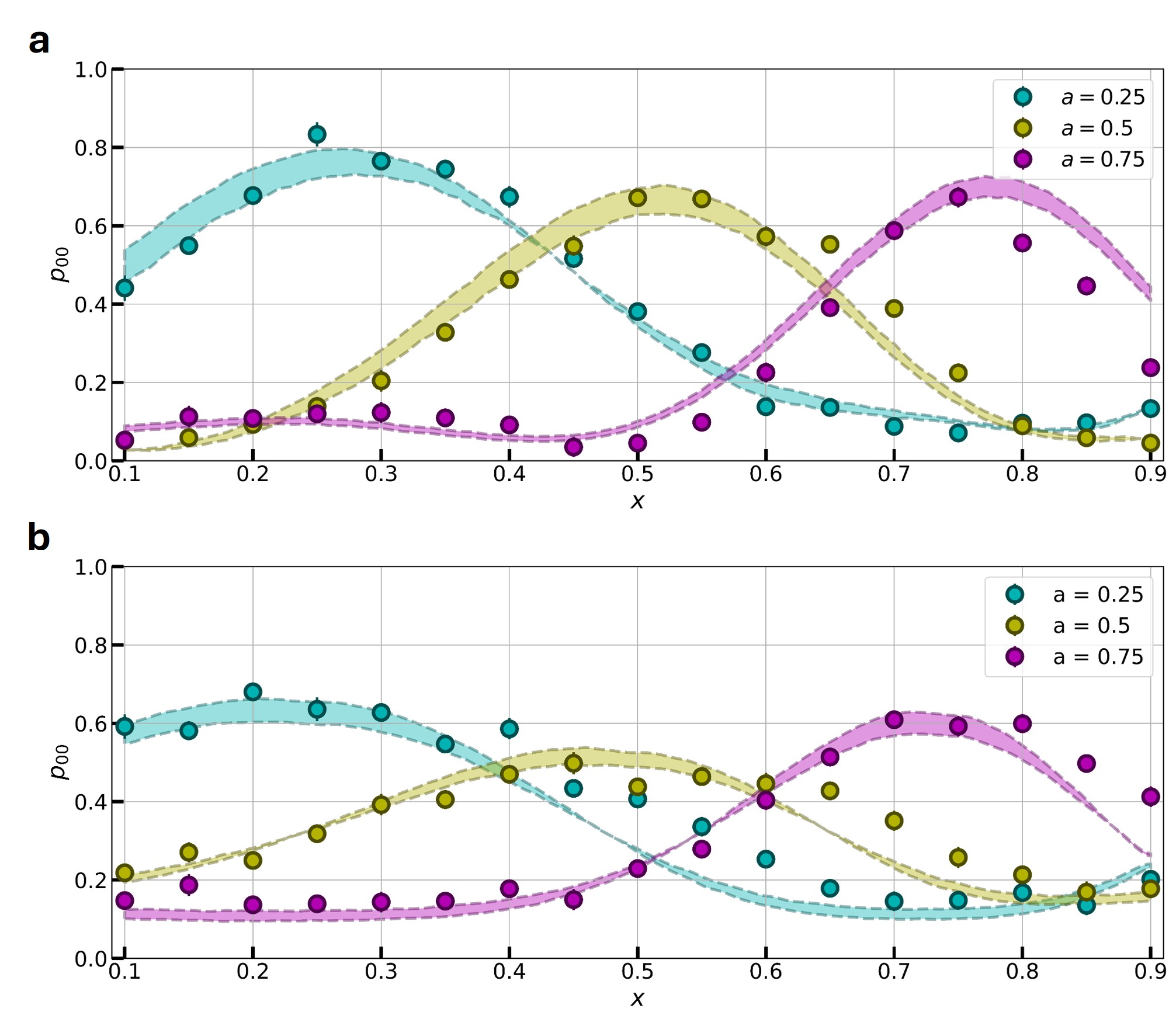}
\caption{
    \textbf{Error modelling of the quantum kernel circuits. a.} Simulated curves of the quantum kernel using our error model, and comparison with the experimental data points. The shaded area corresponds to variations of the error parameters within a $\pm 10\%$ range. \textbf{b.} Simulated curves of the logical quantum kernel using our error model, and comparison with the experimental data points. The shaded area corresponds to variation of the error parameters within a $\pm 10\%$ range. \label{fig:error_model}
}
\end{figure}

\subsection{Comparison between error model and experimental results} 

In order to better understand our results, and especially the source of the distortions of the physical quantum kernel, we run density-matrix simulations of our circuits, using a circuit-level noise model.
The results of the simulations for the physical and logical quantum kernels are compared to the experimental results in Figure \ref{fig:error_model}a and \ref{fig:error_model}b respectively. 
The shaded area corresponds to a variation of the error model parameters within a meaningful range for our system, which is of about $\pm 10\%$ in practice. 
We observe a good agreement between the experimental data and the error model, both for the physical and the logical quantum kernels. Some finite discrepancies between the experimental data and the simulation results remain, which may be explained by experimental fluctuations, imperfect error parameter estimations, and error mechanisms that may not be captured by the error model. We leave further refinement of the error model to future work. 

We find that errors during the initialization, atom transport and measurement have a negligible impact on the outcome with respect to other errors. The most impactful errors are: 
\begin{enumerate}[noitemsep]
    \item Single-qubit gates errors: depolarizing noise and a systematic under- or over- rotation. 
    \item Two-qubit gates errors: Pauli errors, atom loss and a systematic single-qubit phase error due to the imperfect calibration of the compensation $Z$ gate for the accumulated single atom phase.
    \item Local $Z$ gates errors: phase flip errors and atom loss.
\end{enumerate}

We study the effect of each type of noise on the quantum kernel after running the physical circuit, by removing all the other error channels. We observe that: 
\begin{itemize}
    \item All loss-type error channels reduce the contrast, while the minima of the curves remain at zero.
    \item Pauli errors lead to a reduced overall contrast and the appearance of bumps at $x=0.75$ and $x=0.25$ in the $a=0.25$ and $a=0.75$ curves respectively.
    \item Over(under)-rotations during global single-qubit gates reduce the contrast of all three curves, induce unwanted bumps, and shift the maxima of the $a=0.25$ and $a=0.75$ curves toward higher (lower) and lower (higher) $x$-values, respectively.
    \item The systematic single-qubit phase error after CZ gates also reduces contrast, and introduces distortions.
    \item Combining the last two induces extra distortions, shifting the maximum of the $a=0.5$ curve to higher $x$ values.
\end{itemize}
The quality of the quantum kernel is thus mainly degraded because of Pauli errors and coherent errors. Atom loss reduces contrast of the curves, but since the position of the maximum remains identical and the curves do not undergo further distortions, it has a significantly lower impact.

\begin{figure}[t!]
\centering
\includegraphics[width=86mm]{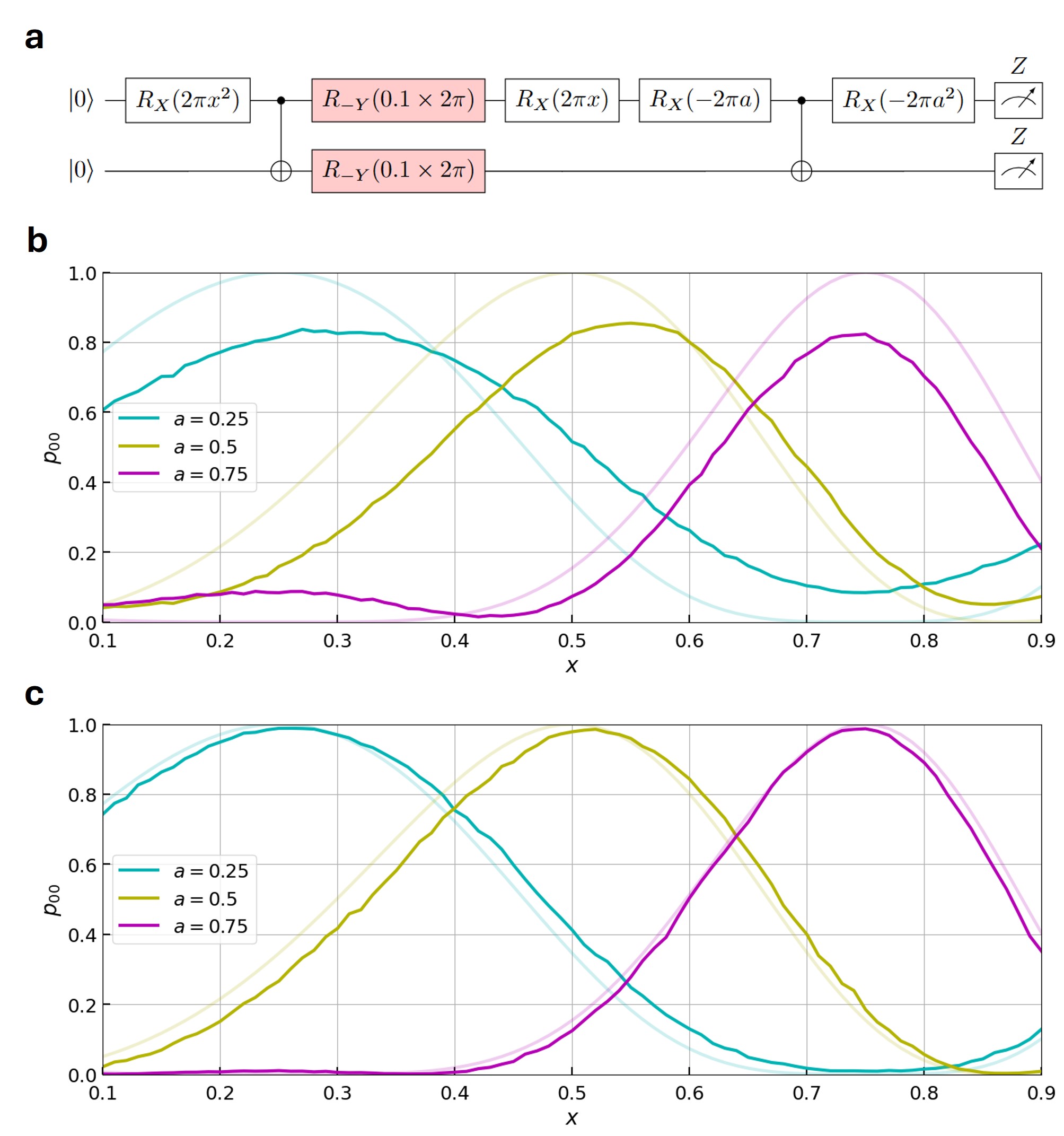}
\caption{
    \textbf{Noise injection at the physical and at the logical level. a.} 
    Quantum kernel circuit with an injected $R_{-Y}(0.1\times2\pi)$ to simulate an error. \textbf{b.} Obtained physical quantum kernel curves with the noise injection. The contrast is reduced and deformations appear: for instance, the maximum of the $a=0.25$ and $a=0.5$ curves are shifted to higher $x$ values, the maximum of the $a=0.75$ curve is shifted to lower $x$ values, and a bump appears at $x=0.25$ in the $a=0.75$ curve. \textbf{c.} Obtained logical quantum kernel curves with the noise injection. The noise is mitigated  at the logical level and minimal deformations appear compared to the physical quantum kernel. \label{fig:noise_injection}
}
\end{figure}

\bigskip
\subsection{Injecting noise in the quantum circuits}

In order to show that kernel distortions are induced by coherent errors and that the logical encoding is able to catch these errors and restore the kernel shape, we inject noise in the form of extra rotations in the physical circuit and check the impact on the resulting curves, without any other type of noise or errors. We then compare with the behaviour of the logical quantum kernel when the same error is injected at an equivalent moment of the circuit.

As an example, we consider a coherent rotation of $0.63\,\text{rad}$ around the $-Y$ axis (Figure~\ref{fig:noise_injection}a). This error, which contains both $X$ and $Z$ components, is representative of typical over-rotations in global single-qubit gates and therefore provides a relevant test case for understanding their impact.
In the physical circuit, such an error reproduces the qualitative features observed in Figure \ref{fig:fig3}e: reduced contrast across all curves, a shift of the $a=0.5$ maximum towards higher $x$ values, and a characteristic “bump” around $x=0.25$ in the $a=0.75$ curve, as shown in Figure~\ref{fig:noise_injection}b.
At the logical level, however, these distortions are strongly suppressed. When mapped onto the logical circuit, the same error does not significantly alter the kernel structure, as illustrated in Figure~\ref{fig:noise_injection}c. Only a mild shift of the $a=0.5$ maximum remains, consistent with experimental trends.
This difference originates from the error-detection structure of the code. The injected error is either flagged by the ancilla qubits or leads to outcomes outside the codespace, which are removed by a parity check enforcing code consistency. In this setting, $17(1)\%$ of the shots are rejected by the ancillas, followed by an additional $17(1)\%$ rejection from the parity check, resulting in a total discard rate of $32(2)\%$.
The resulting effect is an effective filtering of faulty trajectories, which explains the strong suppression of kernel deformations at the logical level. Although the precise distortion pattern depends on where the error is introduced in the circuit, in all tested configurations, introducing an error of comparable magnitude in the logical circuit leads to rejection rates on the order of $20\%$ to $30\%$, indicating that the dominant effect of the code is to identify and discard faulty runs rather than allowing a distortion of the logical kernel.

\end{document}